\documentclass[acmlarge]{acmart}

\makeatletter
\newcommand{\confshort}{\acmConference@shortname}
\newcommand{\conffull}{\acmConference@name}
\newcommand{\confdate}{\acmConference@date}
\newcommand{\confloc}{\acmConference@venue}
\AtBeginDocument{
  \fancypagestyle{firstpagestyle}{
    \fancyhead{}%
    \fancyfoot[C]{}%
  }
  \fancyhf{}
  \fancyhead[LO]{\@headfootfont\shorttitle}%
  \fancyhead[RE]{\@headfootfont\@shortauthors}%
  \fancyhead[LE]{\@headfootfont\footnotesize \confshort, \confdate, \confloc}%
  \fancyhead[RO]{\@headfootfont\footnotesize \confshort, \confdate, \confloc}%
  \fancyfoot[C]{}%
}
\makeatother
\acmBooktitle{\conffull\@ (\confshort), \confdate, \confloc}

\AtBeginDocument{%
  }

\copyrightyear{2026}
\acmYear{2026}
\setcopyright{cc}
\setcctype{by-nc-nd}
\acmConference[FAccT '26]{The 2026 ACM Conference on Fairness, Accountability, and Transparency}{June 25--28, 2026}{Montreal, QC, Canada}
\acmBooktitle{The 2026 ACM Conference on Fairness, Accountability, and Transparency (FAccT '26), June 25--28, 2026, Montreal, QC, Canada}
\acmDOI{10.1145/3805689.3806465}
\acmISBN{979-8-4007-2596-8/2026/06}

\usepackage{booktabs}
\usepackage{subcaption}
\usepackage[table]{xcolor}
\usepackage{array}
\usepackage{csquotes}
\usepackage{comment}
\usepackage{cleveref}
\usepackage{quoting} 
\quotingsetup{vskip=2pt}
\crefformat{section}{\S#2#1#3}
\crefformat{subsection}{\S#2#1#3}
\crefformat{subsubsection}{\S#2#1#3}
\usepackage{pifont}

\usepackage{tikz}

\makeatletter
\newcommand*{\radiobutton}{%
  \@ifstar{\@radiobutton0}{\@radiobutton1}%
}
\newcommand*{\@radiobutton}[1]{%
  \begin{tikzpicture}
    \pgfmathsetlengthmacro\radius{height("X")/2}
    \draw[radius=\radius] circle;
    \ifcase#1 \fill[radius=.6*\radius] circle;\fi
  \end{tikzpicture}%
}
\makeatother

\definecolor{highestcolor}{HTML}{DC143C}
\definecolor{secondcolor}{HTML}{FF8C00}

\begin{document}

\title{\emph{Making a Name for Myself}: On Academic Naming Policies and their Impact}

\author{A Pranav}
\orcid{0000-0001-8951-1517}
\affiliation{%
  \institution{Trustworthy AI Lab, University of Hamburg}
  \country{Germany}}

\author{Vagrant Gautam}
\orcid{0000-0002-7263-8578}
\affiliation{%
  \institution{Heidelberg Institute for Theoretical Studies}
  \country{Germany}}

\author{Martin Mundt}
\authornote{Equal role in advising.}
\orcid{0000-0003-1639-8255}
\affiliation{%
  \institution{University of Bremen}
  \country{Germany}}

\author{Jordan Taylor}
\authornotemark[1]
\orcid{0000-0002-0896-992X}
\affiliation{%
  \institution{Carnegie Mellon University}
  \country{USA}}

\author{Arjun Subramonian}
\authornotemark[1]
\orcid{0000-0002-0415-3800}
\affiliation{%
  \institution{Queer in AI}
  \country{USA}}

\author{Franziska Sofia Hafner}
\authornotemark[1]
\orcid{0000-0003-1070-8267}
\affiliation{%
  \institution{University of Oxford}
  \country{United Kingdom}}

\author{Daniel Chechelnitsky}
\authornotemark[1]
\orcid{0009-0005-6559-254X}
\affiliation{%
  \institution{Carnegie Mellon University}
  \country{USA}}

\author{William Agnew}
\authornotemark[1]
\orcid{0000-0002-1362-554X}
\affiliation{%
  \institution{Carnegie Mellon University}
  \country{USA}}

\author{Anne Lauscher}
\authornotemark[1]
\orcid{0000-0001-8590-9827}
\affiliation{%
  \institution{Trustworthy AI Lab, University of Hamburg}
  \country{Germany}}

\begin{abstract}
In academic publishing, names connect scholars to their work.
When scholars change their names, including for marriage, academic recognition, or gender transition, they may lose credit for past publications.
However, despite significant impacts on citation accuracy and researcher well-being, no existing studies examine how naming policies in computer science serve researchers who change their names.
We use a mixed-methods approach combining surveys ($N=36$), interviews ($N=11$), and large-scale citation analysis of papers from eight major computer science venues from 2019--2025.
We document the multi-year advocacy effort that established the first name change policies, identify implementation barriers including incomplete publisher updates and months-long processing delays.
Researchers continue being cited with misparsed and incorrect names despite publisher updates.
When these citation errors happen, interviewees report significant mental health impacts, including stress, anxiety, and safety risks.
Empirically, we find that venues with accessible and visible name change policies have significantly fewer citation errors compared to inaccessible policies (899 vs. 996 errors per 1,000 papers; $p < 0.001$).
Our annotation analysis shows that deadnaming of transgender researchers in citations decreased by 92\% from 2019 to 2024.
Our findings demonstrate the importance of inclusive publishing policies, for which name change policy advocacy led by trans researchers has been a significant driver.
We recommend that venues adopt proactive visible name change policies, support queer advocacy groups, and improve publication infrastructure to build an inclusive publishing landscape.
\end{abstract}

\ccsdesc[500]{Information systems~Digital libraries and archives}

\keywords{academic publishing, name change policies, scholarly infrastructure}

\renewcommand{\shortauthors}{Pranav et al.}
\maketitle

\section{Introduction}
\label{sec:intro}

\begin{figure*}[t]
\centering
\includegraphics{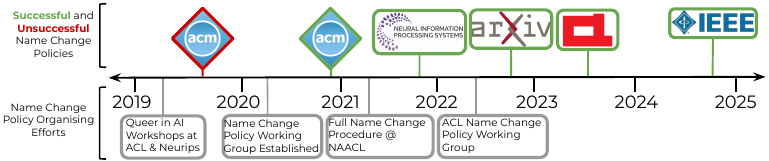}
\caption{A timeline of milestones in name change policies for ACM and ML/NLP venues. Following initial efforts and setbacks and owing to organizing by the Name Change Policy Working Group, Queer in AI, and other organizations, publishers have implemented name change policies.}
\Description{Timeline of name change policy advocacy efforts and venue-level policy adoptions from 2019 to 2025.
The lower track records organizing efforts: Queer in AI workshops at ACL and NeurIPS (2020), the Name Change Policy Working Group established (2021), a full name change procedure at NAACL (2022), and the ACL Name Change Policy Working Group (2023).
The upper track records policy outcomes, distinguishing successful adoptions (green borders) from unsuccessful ones (red borders): ACM's first policy (unsuccessful, 2020), ACM's revised policy (successful, 2021), NeurIPS (2022), arXiv (2023), ACL Anthology (2023), and IEEE (2025).}
\label{fig:ncp_timeline}
\end{figure*}

Personal names serve to uniquely identify individuals as well as categorize them across dimensions of relevance to a society \cite{hough-2016-oxford}. In academic publishing, they serve as fundamental links between scholars and their work. A scholar's record shapes their professional identity, determines how their contributions are recognized and credited, and can shape entire career paths \cite{Jeshion2009TheSO, gautam-etal-2024-stop}. However, rigid systems and hegemonic naming standards often result in scholars publishing under a different name than the one they use in legal contexts, with family, or with friends \cite{Baker,lazet2022case}. 

Contrary to how names are treated as immutable anchors in publishing infrastructure \cite{hunt2020academic}, names can differ for a variety of reasons.
For many researchers, current academic naming conventions cause harm. For example, Chinese names place surnames first, but Western systems routinely reverse them \cite{gasparyan2016scientific}. 
Scholars from geographical regions writing systems outside the global academic norms adopt pen-names or transliterations to conform to present publication standards. South Indian scholars often lack surnames entirely, and publishing forms force them to use their father's given name, creating confusion about who deserves credit for the work \cite{puniamoorthy2008give}.
Patriarchal archival systems have historically erased women's scholarly identities by crediting work under their spouse's name~\cite{anthony2018eradicating}.
These are not edge cases but systematic failures of infrastructure designed for one group (white cis male scholars, with Western first and last names) and imposed on everyone else; systematic failures that further fail to account for the malleable nature of human identity.

In fact, names change throughout researchers' lives, for reasons including marriage, divorce, estrangement, gender transition, public recognizability, parsability for automated systems, privacy, immigration, and religious or cultural reasons \cite{tescione1998research,hough-2016-oxford,Alford_1988,Sue_Telles_2007,Obasi_Mocarski_Holt_Hope_Woodruff_2019}. Yet publishing systems that are built around the assumption that names never change, fail to accommodate this reality.
If a scholar's name change is not updated in their past publications, this leads to a fundamental problem: \textit{Loss of linkage} \cite{hong2004system}. When metadata systems cannot connect old and new names, searches under a new name fail to retrieve publications under a prior name, even though both belong to the same person \cite{bennett2006name}. 
For researchers who change their names, loss of linkage can entail fragmented citation records across databases \cite{pellack2011ripple}, 
due to missing given names, incorrect surnames, bad transliterations, and persistent misspellings \cite{gupta2021missing}. For transgender researchers, on top of these already severe harms, loss of linkage can have even more severe consequences. It results in deadnaming,\footnote{Deadnaming is the act of referring to transgender people by their birth name after they have chosen a new name.} the continued use of prior names that exposes them to discrimination and denies them credit and dignity \cite{vagrant_violence}. Being referenced by a prior name violates the safety and mental well-being of many researchers \cite{Lane_2025}, exposes them to violence \cite{lyght2024sources} and ongoing trauma \cite{TowardsA86:online}, and outs them without consent \cite{tanenbaum2020publishers}. In addition, the inaccessibility of name changes in academia imposes significant labor on scholars who wish to have past publications corrected \cite{Tanenbaum2021DevelopingPractices}.

Although naming in academia carries significant personal and professional weight, we lack a systematic understanding of where academic naming infrastructure helps researchers and where it causes harm.
Prior work has documented individual experiences with name changes and advocated for policy reforms, but no study systematically quantifies how these policies function in practice or measures their impact on citation accuracy.
To address this gap, we ask: \textbf{How do academic naming practices impact researchers, and what harm do they cause?}
We explore this research question through a mixed-methods approach, investigating it both through the lived \textit{experiences} of researchers (\cref{sec:researcher-experiences}) and through \textit{empirical} analysis of naming errors (\cref{sec:citations}).

\textit{1. Experiential analysis.}
We surveyed 36 researchers and interviewed 12 participants who changed their names in academia.
Queer scholars, for whom the stakes are highest, led the grassroots advocacy that established name change policies at ACM and subsequently other venues.
Yet implementation remains inconsistent: Researchers face fragmented citation records that obscure their contributions, and trans researchers additionally experience deadnaming that exposes their identity and compromises their safety.
Errors propagate through reused bibliographies, stale citation manager caches, LLMs that generate hallucinated names, and search engines that lag months behind publisher updates.

\textit{2. Empirical analysis.}
We extracted bibliographies across eight major AI conferences (ACL, EMNLP, NAACL, EACL, NeurIPS, AAAI, FAccT, ICLR, ICML) published between 2019 and 2025.
We compared cited author names against authoritative databases to identify discrepancies, then conducted manual annotation to characterize error types.
We classify a venue's name change policy as \textit{visible} when the proceedings or conference website explicitly describes the option to update author names and links to the process, and \textit{accessible} when a researcher can initiate the change without approval from co-authors, editors, or committees, and without public disclosure of the reason.
By this criterion, ACL Anthology and NeurIPS qualify: Each proceedings volume links directly to the name change form, and both provide author-controlled BibTeX citations.
FAccT falls short because it operates under ACM's policy without mentioning it in the proceedings or on the conference website; AAAI, ICLR, and ICML publish no visible information about name changes.
Our analysis showed that conferences with visible and accessible policies had significantly fewer citation errors than those without ($\chi^2 = 93.97$, $p < 0.001$).
We also found that deadnaming in citations declined following policy adoption: At venues like NeurIPS, rates dropped from 21.5 per 1,000 papers in 2019 to 0.5 per 1,000 in 2024.

Based on these findings, we argue that name change policies correlate with improved citation accuracy, and conclude with three recommendations for inclusive publishing (\cref{sec:recs}): Policy reforms that prioritize author autonomy, technical reforms that include automated citation verification, and sustained community engagement with advocacy groups.

\subsection{Background on Name Change Policies}
Name authority control in library and information science links works by a single author, distinguishes authors with identical names, and supports reliable retrieval across name variants \cite{shiraishi2019accuracy}.
Traditional authority control treats names as externally verifiable facts under institutional control, not as identity claims authors themselves hold \cite{bennett2006name}.
For trans authors, this produces concrete harms: Cataloging practices have surfaced prior names through subject headings without consent \cite{thompson2016more,polebaum2019violent}, and archival systems impose cisnormative structures that do not accommodate identity changes \cite{rawson2009accessing,roberto2011inflexible}.
Queer in AI \cite{Queerinai_2023} and the Queer Metadata Collective \cite{the_queer_metadata_collective_2025_15041154} argue instead that authors should hold authority over their own records.

Trans scholars led sustained advocacy to reform publisher practices: Before 2017, no major publisher had a formal name change policy, and trans authors negotiated corrections case by case while meeting recurring objections that ranged from appeals to the integrity of the historical record to demands for legal proof of a name change \cite{tanenbaum2020publishers}.
A \textbf{name change policy} sets out how authors update their names on already-published works.
An inclusive policy permits the change \textbf{without legal documentation}, applies the update \textbf{across PDFs, metadata, and citation indexes}, and avoids \textbf{errata or other public juxtaposition} of prior and current names \cite{tanenbaum2020publishers}.
\citet{tanenbaum2021vision} formalize these as five principles: accessibility, comprehensiveness, invisibility, expediency, and systematic maintenance.
The Name Change Policy Working Group (NCPWG) and Queer in AI extended this advocacy across publishers, engaged the Committee on Publication Ethics on inclusive-policy guidance \cite{COPE2024}, and contributed to policy adoption at major AI conferences \cite{zhu-2025-standardizing}.

Adoption has not produced uniform implementation.
Some publishers still require legal documentation \cite{watson2023our}, while others cause harm through partial updates or errata that expose prior names \cite{yon2024rules}.
These policies also do not reach errors in citations, which propagate through reused bibliographies and through citation managers such as Mendeley and Zotero that additionally restrict edits at a technical level \cite{the_queer_metadata_collective_2025_15041154}.
This paper documents the struggles researchers face under current naming infrastructure and provides empirical evidence that name change policies improve citation accuracy.
In the upcoming sections, we explore how academic naming practices impact researchers and what harm they cause.

\section{Researcher Experiences with Name Changes and Naming Issues}
\label{sec:researcher-experiences}

Academic publishing systems create barriers which affect researchers who change their names due to gender transition or marriage, as well as those whose names include particles, multiple components, or non-Latin characters.
To understand these experiences, we conducted a survey and semi-structured interviews with affected researchers.

\subsection{Methods}

\begin{table*}[t]
\caption{Overview of interview participants, publisher processing times according to interview participants, and survey findings}
\label{tab:combined}
\begin{minipage}[t]{0.48\textwidth}
\centering
\subcaption{Interview Participants ($N=11$)}
\label{tab:participants}
\small
\begin{tabular}{@{}ll@{}}
\toprule
\textbf{ID} & \textbf{Context} \\
\midrule
P1 & Gender transition \\
P2 & Gender transition; Policymaker \\
P3 & Gender transition; Policymaker \\
P5 & Gender transition \\
P6 & Gender transition; Policymaker \\
P7 & Non-Normative name (East Asian) \\
P8 & Non-Normative name (Latin American) \\
P9 & Non-Normative name (Arabic) \\
P10 & Gender transition; Policymaker \\
P11 & Marriage; German name particles \\
P12 & Gender transition; Policymaker \\
\bottomrule
\end{tabular}

\vspace{1em}

\subcaption{Publisher Name Change Processing Times}
\label{tab:publisher-timing}
\small
\begin{tabular}{@{}llr@{}}
\toprule
\textbf{Publisher} & \textbf{Time} & \textbf{N} \\
\midrule
\multicolumn{3}{l}{\textit{Fast ($<$ 1 month)}} \\
Sci-Hub & Immediate & 1 \\
OpenReview & 5--7 days & 1 \\
ACL Anthology & 2 weeks & 2 \\
CVPR (IEEE) & 2 weeks & 1 \\
arXiv & 2--3 weeks & 1 \\
\midrule
\multicolumn{3}{l}{\textit{Slow (3+ months)}} \\
ACM Digital Library & 6--9 months & 4 \\
Elsevier journals & 6--9 months & 1 \\
PLOS One & 2+ years & 1 \\
\midrule
\multicolumn{3}{l}{\textit{Ongoing/Unresolved}} \\
\textbf{Google Scholar} & \textbf{1+ year} & \textbf{13} \\
SciVal/Elsevier & Ongoing & 1 \\
Springer corrections & Ongoing & 1 \\
\bottomrule
\end{tabular}
\end{minipage}%
\hfill
\begin{minipage}[t]{0.48\textwidth}
\centering
\subcaption{Survey Findings ($N=36$)}
\label{tab:survey}
\small
\begin{tabular}{@{}lr@{}}
\toprule
\textbf{Reasons for name change} & \textbf{N} \\
\midrule
\quad Gender transition & 24 \\ 
\quad Non-Western naming conventions & 5 \\ 
\quad Marriage & 3 \\ 
\quad Academic visibility & 2 \\ 
\midrule
\textbf{Finding} & \textbf{N} \\
\midrule
\multicolumn{2}{@{}l}{\textit{Name components changed}} \\
\quad First name & 19 \\ 
\quad Middle name & 11 \\ 
\quad Last name & 10 \\ 
\quad Full name & 3 \\ 
\quad Particles or diacritics & 3 \\ 
\midrule
\multicolumn{2}{@{}l}{\textit{Citation practices}} \\
\quad Google Scholar as primary source & 22 \\ 
\quad Reuse bibliographies without verification & 5 \\ 
\midrule
\multicolumn{2}{@{}l}{\textit{Old name persistence ($N=22$ tracked)}} \\
\quad Still cited with old name & 18 \\ 
\midrule
\multicolumn{2}{@{}l}{\textit{Publisher update outcomes}} \\
\quad Partial updates only & 10 \\ 
\quad Complete removal achieved & 4 \\ 
\midrule
\multicolumn{2}{@{}l}{\textit{Reported impacts}} \\
\quad Significant stress or anxiety & 12 \\ 
\quad Fragmented citation metrics & 11 \\ 
\quad Unwanted disclosure of identity & 10 \\ 
\quad Affected sense of belonging & 8 \\ 
\quad Reported no impacts & 8 \\ 
\midrule
\multicolumn{2}{@{}l}{\textit{Policy awareness}} \\
\quad Unsure if publishers have policies & 18 \\ 
\quad Clear, well-implemented policies exist & 1 \\ 
\bottomrule
\end{tabular}
\end{minipage}
\end{table*}

\paragraph{Survey.}
We distributed an anonymous survey through academic and personal networks on BlueSky (\url{bsky.app}) and at multiple institutions, as well as through Queer in AI channels, between October and November 2025.
Respondents ($N=36$) answered questions about their experiences with name changes, naming issues, citation practices, and publisher interactions.
All questions were optional to reduce participant burden on sensitive topics, and participants could request deletion of their data at any time.
The survey captured researchers across disciplines including natural language processing, machine learning, robotics, and linguistics.
The full survey is provided in Appendix~\ref{app:survey}.

\paragraph{Limitations.}
Recruitment is difficult because name changes are sensitive and often undisclosed.
We advertised through academic networks beyond trans-specific communities, but most respondents were trans researchers (58.3\% of survey respondents and 7 of 11 interviewees).
Their experiences ground our analysis because trans scholars built these policies and bear the highest stakes from deadnaming, including outing and harassment.
We treat the survey ($N=36$) as thematic evidence alongside the interviews, not a basis for statistical generalization.

\paragraph{Interviews.}
We contacted 11 scholars (survey respondents and Name Change Policy Working Group members who expressed interest) for semi-structured interviews, to better understand their name change history and motivations, interactions with publishers and databases, technical challenges encountered, involvement in policy advocacy (where applicable), and emotional and professional impacts.
We sought to represent three categories of participants: Scholars who changed names due to gender transition or marriage, scholars navigating non-normative naming conventions, and researchers involved in developing name change policies.
All questions are shown in Appendix~\ref{app:interview}, but we allowed participants to skip questions and guide the depth of the conversation.
Interviews were conducted via video call between September and December 2025, and lasted $30$--$45$ minutes each.
With participant permission, we recorded and transcribed all interviews for analysis.
We refer to participants as P1--P12 (Table~\ref{tab:participants}) to preserve anonymity, as the name change advocacy community is small and participants may be identifiable from detailed descriptions.

After transcribing each interview, the first and second author independently open-coded the transcripts \cite{corbin2014basics}. Over three weeks, these authors met to discuss shared patterns across the interview codes and survey responses. Based on these, we identified common sociotechnical barriers to name changes, such as venue policies and technological publication infrastructure. Additionally, we explored how these barriers shape name change decisions and the consequences of deadnaming.
Overall, we identify 5 interrelated themes in researcher experiences with name changes: (1) How deadnaming harms researchers, (2) how publication infrastructures perpetuate incorrect naming, (3) how incorrect naming propagates through citations, (4) how researchers make naming choices, and (5) how researchers mitigate deadnaming through name change policy advocacy. We organize the findings below around these themes, integrating survey data where quantitative patterns corroborate interview accounts.
Table~\ref{tab:combined}c summarizes key findings.


\subsection{How Deadnaming Harms Researchers}

We asked participants how name changes affected their wellbeing and careers.
Their responses revealed that the process extends beyond administrative inconvenience into significant emotional, psychological, and safety-related harm.

\subsubsection{Changing names and deadnaming causes trauma and stress.}

Two participants (P2, P6) described severe psychological consequences. P6 experienced dissociative amnesia when navigating bureaucratic processes, losing track of name change requests because she was not forming working memories of those interactions. Six years post-transition, she reported visceral negative reactions to their former name. P2 expressed the disappointment that she faced when dealing with resistances regarding name changes in academia:
\begin{quoting}
``\textit{It just felt like the worst kind of irony that every institution in society was supporting me except for mine. And that was painful, because I've really built a lot of identity and investment in the institution of higher education. And to have this thing that I'd spent so much time investing in and helping build---and now as a senior faculty member, helping to maintain and sustain---that institution being the one that couldn't get its act together, hurt me. And it still continues to be the worst of all of those settings for name changes.}''
\end{quoting}
33.3\% of the authors who have changed the names reported stress or anxiety from the name change process, 30.6\% experienced fragmented citation metrics, 27.8\% experienced unwanted disclosure, and 22.2\%  reported effects on their sense of belonging.
One respondent expressed \textbf{fear about future submissions to venues requiring legal names.}

\subsubsection{Badly implemented policies create safety risks.}
Inadequate name change policies create serious safety concerns for trans researchers. When publishers fail to discreetly handle name changes, they risk outing researchers who have not publicly disclosed their trans identity.
Two participants (P2, P3) described documented cases where trans researchers faced violence after being outed through academic records.
Publisher records often contain affiliation information that can reveal a researcher's location to those who might wish them harm. 
P6 described their experience with Springer's incomplete updates:
\begin{quoting}
``\textit{Since my affiliation information is included on each paper, Springer has now disclosed my location to anyone who might wish me harm simply for being a transgender woman. You might as well have just put up a big red tag: WARNING THIS AUTHOR IS TRANSGENDER! These concerns are not hypothetical, nor are they an academic exercise for me: They are a matter of personal safety. }''
\end{quoting}
P6 further attributed that that poor name change policies amplify surveillance risks by creating additional paper trails that link old and new identities.

\subsubsection{Participants expressed concern about policy rollbacks amid the growing anti-DEI political climate.}
Two participants (P3, P6) expressed concern about potential policy rollbacks amid waves of anti-trans political climates and growing anti-diversity movements.
These concerns are already materializing: Queer in AI has lost sponsors due to the current U.S. political climate~\cite{queerinai:personalcomm2025}.
P3 worried that publishers might remove existing name change protections. P6 observed that public-facing equity initiatives have disappeared at some institutions.


\subsection{How Publication Infrastructures Perpetuate Incorrect Names}

We asked participants how long publishers took to handle name change requests and how completely records were updated.
Their responses revealed that even when policies exist, implementation remains problematic.

\subsubsection{Incomplete updates to name changes cause harm.}

Among survey respondents who attempted name changes, only 11.1\% ($N=4$) achieved complete removal of their deadname. The majority reported partial fixes: 30.6\% reported updates where some records changed while others remained untouched.
These partial updates leave deadnames visible in parts of the publication record.
A publisher may change metadata but leave the deadname in the PDF or may update the website but leave the deadname in citation exports \cite{TowardsA86:online}.
The result is a ``fixed'' record that still exposes the author's deadname and propagates it through downstream systems.
P10 described that citation managers pull unchanged metadata, colleagues download old PDFs, and deadnames propagate through the very system that was supposed to stop it. P6 described their experience with Springer's incomplete updates: \textit{``Whether Springer intends it to be or not, this is a form of discrimination that amounts to a misattribution of my scholarship.''}

\subsubsection{Name updates range from weeks to months.}

Table~\ref{tab:publisher-timing} summarizes reported processing times across publishers. Among our interviewees, four publishers responded in days: OpenReview processed requests in 5--7 days, ACL Anthology and IEEE venues took 2 weeks, and arXiv took 2--3 weeks. 
Six publishers required months. Cambridge University Press took several months. ACS took 3--4 months. Elsevier, Taylor \& Francis, and ACM each required 6--9 months. PLOS One took over two years for one participant. Although reported times are based on select scholars, we posit that it is unlikely that all these values are ``accidental edge cases,'' but rather, they are indicative of a broader systemic problem. 
These timelines subjected scholars to extended periods of corrective labor: P2 and P6 described sending numerous emails to editors and collaborators to push name changes through.
P2 leveraged completed government name changes as negotiating tools to pressure publishers into processing their requests

Scholarly databases have different responses to name change requests. Three participants independently contacted Sci-Hub (\url{www.sci-hub.in}) and received immediate assistance; one participant reported that Sci-Hub accepted an updated PDF with no questions asked. ProQuest is at the opposite end of the spectrum: No name change policy exists, and the per-paper fee makes corrections prohibitively expensive. 
P3 described how faculty hiring processes rely on Scopus metrics, yet only 5 of her publications appeared under her correct name in Scopus.
Fixing this required developer intervention and took several months.

\subsubsection{Google Scholar perpetuates deadnames.}

Google Scholar is a scholarly search engine that crawls academic websites and indexes what is automatically deemed scholarly content.
With no editorial staff, there is no one to email and no policy to invoke for scholars looking to correct their name.
P6 described it as \emph{``an opaque behemoth that does a lot of harm.''}
Thirteen survey respondents reported ongoing issues lasting over a year with correcting their name on Google Scholar.
Our survey also found that 61.1\% of respondents use Google Scholar as their primary citation source.
While researchers can edit how papers appear on their own profile, they cannot change it elsewhere.
P5 explained: \emph{``You can only edit your version of them. You can't edit your co-authors' versions.''}
Google Scholar's paper version feature magnifies this problem.  
Anyone can view other versions of the paper, revealing naming differences to curious or malicious readers. For trans researchers, the platform functions less as a discovery tool than a permanent record of who they used to be.
Queer in AI publicly called on Google Scholar to address these issues in 2022; the company never responded~\cite{TransRes41:online}.


\subsection{How Incorrect Names Propagate Through Citations}

We asked participants whether publisher updates propagated to subsequent citations of their work.
Their responses revealed that incorrect naming and deadnaming persisted even after successful policy outcomes.

\subsubsection{Citation databases propagate incorrect names.}
Google Scholar dominates citation workflows, with 61.1\% ($N=22$) of survey respondents using it as their primary source. An additional 13.9\% ($N=5$) reused bibliographies without verification, propagating deadnames through copy-paste workflows. These practices created cascading errors: Among the 22 respondents who tracked citations, 81.8\% ($N=18$) reported continued incorrect names in recent publications even after publishers updated records.
P2 described creating a distributed advocacy network where colleagues flagged their deadname during peer review to avoid the circulation of deadnames.

\subsubsection{LLM-generated citations have deadnames.}
Researchers increasingly use large language models to generate bibliography entries, but these systems are often trained on historical data that predate name changes \cite{liao2024llms}. 
Prior work has found that GPT-3.5 fabricates up to 55\% of citations and GPT-4 up to 18\%~\cite{Walters2023FabricatedGPTCitations}.
One interview participant explained the scale problem: \textit{``15,000 citations that deadname me and 100 papers that don't---that just swamps all of the training data.''} P6 discovered in the OpenAI lawsuit data dump that all 22 instances of their work used their deadname; P3 similarly found that papers in training datasets were ``all in my deadname.'' After years of declining deadname citations as publishers updated their records, P2 observed a reversal due to more researchers using LLMs for citations: \textit{``Now I've started to see an increase [in deadnames] again.''}


\subsection{How Researchers Make Naming Choices to Fit With Citation Norms}
\label{sec:name-selection}

Researchers change their names for many reasons.
Among our survey respondents, gender transition was the most common reason (58.3\%), followed by marriage (19.4\%), cultural or religious considerations (8.3\%), academic visibility (5.6\%), immigration (5.6\%), and family reasons (5.6\%). 
Privacy, safety concerns, and name formatting issues each accounted for 2.8\%. 
Respondents most often changed first names (52.8\%), followed by middle names (30.6\%) and last names (27.8\%). 
Full name changes occurred in 8.3\% of cases, while 8.3\% adjusted cultural elements like particles or diacritics.
The names researchers adopted reflected both personal meaning and strategic calculation.
P7 and P8 both highlighted the importance of a name in academia for building reputation and brand.
Therefore, technical infrastructure constraints, citation system limitations, and database assumptions all shape how researchers constructed their professional identities.
Below we examine patterns in naming choices.

\subsubsection{Many trans researchers chose their names to preserve initials in citations.}

Among respondents who changed names due to gender transition ($N=21$), first names changed most frequently (90.5\%) as observed in prior work \cite{Obasi_Mocarski_Holt_Hope_Woodruff_2019}. Middle names changed for 38.1\%, full names for 14.3\%, and last names for only 9.5\%. Many trans scholars, including P2 and P5, strategically preserved initials when selecting new names for citation continuity.
P10 and P6 drew on family connections: P10 used a last name as a first name and adopted a maternal surname; P6 took a variant of a family name that another relative had discarded. 

\subsubsection{Participants stressed the importance of choosing names early in their careers.}

Name changes due to marriage involve different calculations. P11 wanted a single consistent family name after getting married and weighed short-term disruption against decades of future publishing: ``\emph{My career is going to be 30--35 years of publishing and I've published for like 5 years so I'd rather have a hard time for 5 years rather than do it later.}'' 
P5 and P7 shared this forward-looking reasoning and expressed relief at having changed their names early, accepting temporary friction to avoid larger problems later in their careers.

\subsubsection{Researchers faced harm when their names do not fit the citation norms.}

Academic naming systems assume a simple first-last structure. This often creates problems for researchers whose names include articles, prepositions, or multiple family names.
P9 has an Arabic name where an article and subject are typically connected. Her passport separates the Anglicized version of her name into two parts, causing only part of her name to appear in citations:
\begin{quoting}
\textit{``I hate that nobody discussed these things or realized the implications\ldots Why is this so hard for me? Why is it not as straightforward as other names\ldots It's extra effort that we have to put in seeking out this knowledge which is not widely available. \ldots This should be discussed in the research group, publishers should say stuff about this, the advisor should discuss this with students, otherwise people will not know this.''}
\end{quoting}
P7, on the other hand, was prompted by a co-author to think about their publication name before submitting their first paper.
She has a common East Asian last name, and flipped the order: ``\emph{I wanted to be more distinctive. If you click on any author name, it pulls up all the papers written by the last name and the first initial. If I stuck with my original name, then arXiv just wouldn't work.}''
P8's Spanish surname is composed of her father's family name (which includes a preposition) and her mother's family name.
Although other academics in the family use various strategies (including dropping the longer father's family name, dropping diacritics or the preposition, and so on), she ``\emph{didn't want any of these options because it's not me.}''
Instead, she chose to hyphenate their father's family name and remove her mother's surname, which felt more authentic and would cause fewer problems with citations.
P11 did not hyphenate her last name, which contains German name particles, and faced challenges.
The ACL Anthology inconsistently handled the capitalization and inclusion of particles in her name, mixing up variations and dropping them in citations.
She had to contact administrators to merge multiple profiles that the system had created for her.
Survey respondents described similar challenges. One noted that adding a hyphen to their last name ``\emph{was resolved on my personal page, but not on the actual publication,}'' resulting in ``\emph{publications with two different last names.}'' Another described constantly needing to ``\emph{double check everything to make sure all three names are visible}'' and ensure their name is not ``\emph{getting misparsed.}''


\subsection{How Researchers Mitigate Deadnaming Through Name Change Policy Advocacy}

We asked participants about their experiences advocating for name change policies.
Their responses documented the formation of the first institutional name change policy in academic publishing, the resistance advocates faced, and the multi-year effort that transformed publisher practices industry-wide.
\textbf{Name change policy advocacy was and continues to be led by trans scholars, as the stakes for trans researchers are fundamentally higher.}

\subsubsection{Researchers reported that they were unaware about name change policies.}
Researchers often lack basic information about name change options. 50\% ($N=18$) of survey respondents were unsure whether publishers in their field have formal policies, and only one person reported clear, well-implemented processes. Before the formation of advocacy networks in 2020, information was scarce; one participant who changed their name in 2016 recalled guessing at available options and disclosing extensive information to publishers, unaware that others had navigated this process before. P1 learned about name change processes through Queer in AI, illustrating how knowledge often may concentrate within advocacy communities rather than official channels.
Three cisgender participants (P8, P7, and P11) reported learning about name change processes from trans colleagues or trans name change advocacy.

\subsubsection{Collapse of the initial efforts of establishing name change policy (2017--2018).}
Before 2017, no major academic publisher had a formal name change policy. Trans scholars who needed to change their names had no process to follow, and simply had to guess what might work. One early advocate described the approach: \emph{``I just bothered the ACM Digital Library people until they made it happen.''} This individual initiated a working group with ACM in 2017--2018, becoming one of the first to push for institutional change. The effort stalled after the advocate received harassment from a senior ACM member, severe enough that ACM had to intervene.
The initiative subsequently collapsed.

\subsubsection{Establishing the ACM Name Change Policy (2019).}
In June 2019, a new group of advocates was recruited at a conference. P6 recalled being approached because \emph{``all of the trans people on [the committee] just quit.''} Over a single weekend in September 2019, they drafted an initial policy proposal. The reception was immediate and total rejection: \emph{``It was rejected completely, 100\%, like s**t-canned. They did not want to give us any of the things that we asked for.''} Months of negotiation followed. ACM approved a compromise policy in December 2019.
The \textbf{final version required signed letters from co-authors} to shift legal liability, an extraordinary step beyond protections afforded to cisgender authors making routine corrections. One advocate who had been on the committee described an earlier sticking point:
\begin{quoting}
\textit{``The initial proposal that came down would have had the \textbf{ACM require copies of people's driver's licenses} and things to prove that they'd really changed their name and then keep copies of them. I'm not sure if they are still actually doing this, but it was like a source of irritation and distress, like this kind of presumption that \textbf{trans people are in some way fraudulent or deceptive}.''}
\end{quoting}

\subsubsection{ACM name changing queue (2019--2020).}
Implementation of the ACM name change policy created new bottlenecks. ACM processed name changes serially, one researcher at a time. The first applicant received changes within 48 hours. Researchers later in the queue waited months. One senior scholar, third in the queue, waited six to seven months for completion, blocking everyone behind them the entire time. They described the infrastructure: \emph{``ACM just had part-time external contract staff changing PDFs one at a time.''} A fifth applicant waited approximately nine months. This serial process meant that one prolific researcher's backlog determined whether others could have their names corrected at all.
The queue also created leverage. One advocate described how pending name changes were used as a bargaining chip: \emph{``My name change got done with the ACM while the policy was still being discussed, but other committee members have been whipped into obedience by dangling [the name change] in front of them.''}

\subsubsection{Development of future policies and Name Change Policy Working Group (2020--).}
Despite its implementation failures, the ACM policy became a template for other publishers. 
A Discord server was created that year, hosting the practical information necessary to navigate ambiguous publisher websites and identify the right contacts when official documentation remained unclear.
The advocacy effort coalesced into the Name Change Policy Working Group, (\url{ncpwg.org}) which continues to advocate across publishers, including by engaging with the Committee on Publication Ethics (COPE), publishing principles for inclusive name change policies in 2021 (although formal COPE guidance remains pending) \cite{COPE2024}.
Over approximately two and a half years, the group changed the publishing industry, getting every major publisher on board with policies based on their templates \citep{zhu-2025-standardizing}.


\section{Measuring Naming Errors in Citations}
\label{sec:citations}

Academic citations frequently contain incorrect author names, causing fragmented citation metrics and creating stress for affected researchers.
In this section, we present an empirical analysis to understand \textbf{how publication culture (inclusive policies and supporting infrastructure) contributes to naming errors in citations.}
We measure this through automated and manual analysis of naming errors in bibliographies from eight major AI venues.

\subsection{Methodology}
Understanding the scope of citation errors requires systematic measurement across venues and time.
We extract bibliographies of papers across eight computer science venues between 2019 and 2025, compare author names against authoritative databases to identify discrepancies, and manually annotate flagged cases to categorize error types.

\subsubsection{Reference Checking Pipeline}
\label{sec:pipeline}

\begin{figure}[t]
\centering
\includegraphics[width=\columnwidth]{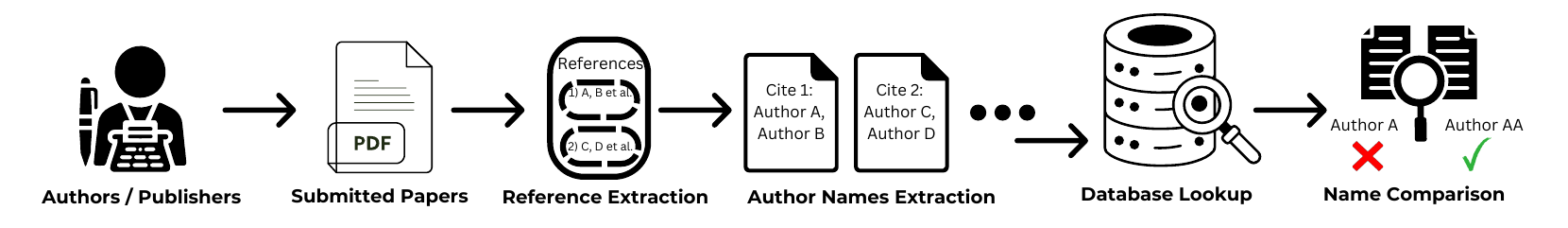}
\caption{The reference checking pipeline: References are extracted from PDFs and author names are validated against authoritative databases. Any inconsistencies are flagged.}
\Description{Pipeline diagram for \textsc{RefCheck} showing six sequential stages connected by arrows.
Authors and publishers submit papers as PDFs.
The reference extraction stage parses the reference list from each PDF.
The author name extraction stage isolates individual author names from each reference.
The database lookup stage queries extracted names against authoritative databases.
The name comparison stage flags discrepancies, illustrated by ``Author A'' matching ``Author AA'' with a red cross and a green checkmark.}
\label{fig:refcheck}
\end{figure}

Measuring citation accuracy at scale requires automated verification infrastructure.
Figure~\ref{fig:refcheck} illustrates our two-stage pipeline: Citation extraction followed by name matching against databases.

\textbf{Citation Extraction.}
We parse references directly from PDFs, as this is the format typically received by publishers during submission.
Our system uses GROBID (GeneRation Of BIbliographic Data), an open-source parser that processes PDFs at 10.6 documents per second~\cite{GROBID}.
The parser extracts four bibliographic fields: Author names, paper titles, venues, and publication years.
We use a BERT-based Conditional Random Field model for parsing~\cite{devlin2019bertpretrainingdeepbidirectional}.
We chose this architecture over LLM-based parsers for computational efficiency, which is necessary for publisher-scale workflows.
We fine-tune the parser on computer science papers, as domain-specific training improves accuracy for citation parsers~\cite{grennan-beel-2020-synthetic}.
For fine-tuning, we randomly sample 3,000 papers from eight major AI conferences (ACL, EMNLP, NAACL/EACL, NeurIPS, AAAI, FAccT, ICLR, ICML) from 2019 to 2025 with \LaTeX{} source files on arXiv.
For each paper, we extract the BibTeX entries and align them with the corresponding references in the compiled PDF, creating parallel training pairs.
We split the dataset into training (80\%), validation (10\%), and test (10\%) sets.
We evaluate using macro-average field-level F1 score, a standard metric for citation parsers~\cite{councill2008parscit}.
Fine-tuning improves F1 from 0.89 to 0.96 on our held-out test set of 300 papers.
Appendix~\ref{sec:parser} provides more detailed information regarding parser construction, and finetuning details like training parameters, evaluation metrics and hardware.

\textbf{Name Matching.}
For each extracted citation, we query three databases: DBLP, ACL Anthology, and arXiv, which we selected because they support author-initiated name updates and are heavily used within AI and NLP.
DBLP also integrates with ORCID, allowing authors to control their own records \cite{dblp_orcid_faq}.
We detect discrepancies using the normalized Damerau-Levenshtein distance to quantify string similarity between the author name and the database entry \cite{Damerau1964}.

\subsubsection{Experimental Setup}

We extract bibliographies from 106,410 papers across eight computer science venues (ACL, EMNLP, NAACL / EACL, NeurIPS, AAAI, FAccT, ICLR, ICML) from 2019 to 2025.
We obtain open-access PDFs through the arXiv API to run through our pipeline \cite{arxivAPI}.
Comparing \textcolor{blue}{extracted names} to \textcolor{teal}{authoritative names} yields eight primary categories of errors described below. We developed this typology inductively by manually reviewing a sample of flagged discrepancies and grouping them by error pattern, and we categorize all detected discrepancies using this typology.

\begin{enumerate}
    \item \textbf{Last name errors (LN-E)}: The cited last name differs substantially from the authoritative record, possibly due to marriage or divorce.
    We classify errors as \textbf{LN-E} when the normalized Damerau--Levenshtein similarity between last names falls below 0.5.
    Example: \textcolor{blue}{Ilya Rozanov} $\rightarrow$ \textcolor{teal}{Ilya Hollander}.
    
    \item \textbf{Typographical errors (Typo)}: Character-level mistakes in either name field as databases or citing bibliographies contain typographical mistakes.
    We classify errors as \textbf{Typo} when similarity falls between 0.5 and 0.9.
    Example: \textcolor{blue}{Nicholas} $\rightarrow$ \textcolor{teal}{Nicolas}.
    
    \item \textbf{Last name compound errors (LN-C)}: Multi-part surnames incorrectly parsed or truncated.
    This error happens when a surname has more than two parts and the parser treats only one of them as the last name.
    We detect these through pattern matching for hyphens, particles (de, van, von), and patronymic components (bin, binti), then verify that the discrepancy appears in the last name.
    Example: \textcolor{blue}{Berg} $\rightarrow$ \textcolor{teal}{van den Berg}; \textcolor{blue}{Smith} $\rightarrow$ \textcolor{teal}{Smith-Johnson}.
    
    \item \textbf{First name compound errors (FN-C)}: Multi-part given names incorrectly parsed or truncated.
    This error happens when a given name has more than two parts and the parser treats only one of them as the first name.
    We apply the same pattern matching as LN-C, including honorific titles (Md.), then verify that the discrepancy appears in the first name field.
    Example: \textcolor{blue}{M. Alexandari} $\rightarrow$ \textcolor{teal}{Amr Alexandari}.
    
    \item \textbf{Nickname errors (Nick)}: Informal or shortened name variants used instead of formal names.
    This error happens when authors are cited by the name they use in everyday life rather than by the name on their publications.
    We classify errors as \textbf{Nick} when the cited name matches a known alias in DBLP's author profile and the similarity between cited and authoritative names exceeds 0.4.
    Example: \textcolor{blue}{Rob} $\rightarrow$ \textcolor{teal}{Robert}; \textcolor{blue}{Bill} $\rightarrow$ \textcolor{teal}{William}.
    
    \item \textbf{Transliteration errors (Trans)}: Mistakes in variant romanizations of non-Latin scripts or diacritical conversions.
    We normalize names using \texttt{unidecode} and custom character mappings (e.g., \"u$\rightarrow$ue, \ss$\rightarrow$ss), then classify as \textbf{Trans} when normalized forms match but raw forms differ.
    Example: \textcolor{blue}{Muller} $\rightarrow$ \textcolor{teal}{M\"uller}.
    
    \item \textbf{Author order errors (AO)}: Authors listed in incorrect sequence or omitted entirely.
    This error happens when the citation parser drops one of the authors from the extracted list.
    We compare author positions in the citation against the authoritative author list.
    Example: citation lists three of four authors, omitting one contributor.
    
    \item \textbf{Wrong person errors (WP)}: Citation attributes work to an entirely different researcher.
    This error happens when researchers share surnames, or when LLM-assisted writing hallucinates citations entirely.
    We classify errors as \textbf{WP} when the cited and authoritative names share the same last name but have different first names in DBLP.
    Example: \textcolor{blue}{Dirk Hovy} $\rightarrow$ \textcolor{teal}{Eduard Hovy}.
\end{enumerate}

\textbf{Metric.}
We report \textbf{errors per 1,000 papers}: For each venue, we divide the raw count of flagged citations by the number of papers analyzed, then multiply by 1,000.
We also compute errors per 1,000 citations and report those results in Appendix \ref{sec:errors}.
We prioritise the per 1000 papers metric because it better reflects harm: A single deadnaming citation causes equal harm whether the citing paper contains 10 or 100 references.

\textbf{Policies.}
Applying the criteria defined in \cref{sec:intro}, we classify ACL, EMNLP, NAACL/EACL, and NeurIPS as venues with visible and accessible policies (see \cref{sec:policy}, Figures \ref{fig:acl} and \ref{fig:neurips}), and AAAI, FAccT, ICLR, and ICML as venues without.
ACL Anthology and NeurIPS additionally provide author-controlled BibTeX citations \cite{bollmann-etal-2023-two}.
Table~\ref{tab:comprehensive-venue-errors} reports error rates per 1,000 papers across three time periods (2019--21, 2022--23, 2024--25).

\begin{table*}[t]
\centering
\caption{Errors in paper bibliographies, per 1,000 papers (2019--2025). 
\textcolor{red}{\textbf{Red}} = highest; \textcolor{orange}{\underline{orange}} = second. Typo = typographical; LN-C/FN-C = last/first name compound; Nick = nickname; LN-E = last name error; Trans = transliteration; AO = author order; WP = wrong person. '19 = 2019--21; '22 = 2022--23; '24 = 2024--25.}
\footnotesize
\renewcommand{\arraystretch}{1.15}
\setlength{\tabcolsep}{2.8pt}
\begin{tabular}{@{}l*{24}{r}@{}}
\toprule
& \multicolumn{3}{c}{\textbf{Typo}} 
& \multicolumn{3}{c}{\textbf{LN-C}} 
& \multicolumn{3}{c}{\textbf{FN-C}} 
& \multicolumn{3}{c}{\textbf{Nick}} 
& \multicolumn{3}{c}{\textbf{LN-E}} 
& \multicolumn{3}{c}{\textbf{Trans}} 
& \multicolumn{3}{c}{\textbf{AO}} 
& \multicolumn{3}{c}{\textbf{WP}} \\
\cmidrule(lr){2-4} \cmidrule(lr){5-7} \cmidrule(lr){8-10} \cmidrule(lr){11-13} 
\cmidrule(lr){14-16} \cmidrule(lr){17-19} \cmidrule(lr){20-22} \cmidrule(lr){23-25}
\textbf{Venue} & \textbf{'19} & \textbf{'22} & \textbf{'24} 
& \textbf{'19} & \textbf{'22} & \textbf{'24} 
& \textbf{'19} & \textbf{'22} & \textbf{'24} 
& \textbf{'19} & \textbf{'22} & \textbf{'24} 
& \textbf{'19} & \textbf{'22} & \textbf{'24} 
& \textbf{'19} & \textbf{'22} & \textbf{'24} 
& \textbf{'19} & \textbf{'22} & \textbf{'24} 
& \textbf{'19} & \textbf{'22} & \textbf{'24} \\
\midrule
\multicolumn{25}{@{}l}{\cellcolor{gray!8}\textit{Venues with visible and accessible name change policies}} \\[2pt]
ACL          & 452 & \textcolor{red}{\textbf{417}} & 435 
             & 106 & 165 & 154 & 93 & 91 & 129 & 134 & 128 & 169 
             & 274 & 315 & 285 & 132 & 82 & 103 
             & 46 & \textcolor{red}{\textbf{57}} & 57 & 41 & 52 & 69 \\
EMNLP        & \textcolor{orange}{\underline{453}} & 377 & \textcolor{orange}{\underline{456}} 
             & 156 & 185 & 178 & 119 & 117 & 115 & 120 & 160 & 132 
             & \textcolor{red}{\textbf{303}} & 299 & 320 & 75 & 86 & 92 
             & 21 & \textcolor{orange}{\underline{50}} & 61 & \textcolor{orange}{\underline{47}} & 44 & \textcolor{orange}{\underline{74}} \\
NAACL/EACL   & \textcolor{red}{\textbf{482}} & \textcolor{orange}{\underline{413}} & 404 
             & 116 & 167 & 185 & 99 & 120 & 143 & 144 & 137 & 153 
             & \textcolor{orange}{\underline{292}} & \textcolor{orange}{\underline{393}} & 321 & 99 & 88 & 85 
             & 21 & 29 & 49 & 28 & \textcolor{red}{\textbf{65}} & 44 \\
NeurIPS      & 298 & 346 & 364 
             & 308 & 305 & 288 & \textcolor{orange}{\underline{228}} & 202 & 204 & 196 & 226 & 200 
             & 168 & 265 & 324 & 111 & 133 & 104 
             & 44 & 49 & 65 & 38 & 54 & 54 \\
\midrule
\multicolumn{25}{@{}l}{\cellcolor{gray!8}\textit{Venues without visible and accessible name change policies}} \\[2pt]
AAAI         & 179 & 139 & 144 
             & 389 & 440 & \textcolor{red}{\textbf{508}} & \textcolor{red}{\textbf{279}} & \textcolor{orange}{\underline{298}} & \textcolor{orange}{\underline{251}} & 217 & \textcolor{orange}{\underline{234}} & 203 
             & 167 & 153 & 213 & 131 & 102 & 95 
             & 43 & 43 & 56 & 37 & 35 & 40 \\
FAccT        & 346 & 360 & \textcolor{red}{\textbf{463}} 
             & \textcolor{red}{\textbf{462}} & \textcolor{red}{\textbf{464}} & 241 & 192 & 186 & 174 & \textcolor{red}{\textbf{308}} & \textcolor{red}{\textbf{284}} & 205 
             & 269 & \textcolor{red}{\textbf{462}} & \textcolor{red}{\textbf{388}} & 115 & \textcolor{orange}{\underline{188}} & \textcolor{red}{\textbf{214}} 
             & \textcolor{red}{\textbf{115}} & 22 & \textcolor{red}{\textbf{93}} & \textcolor{red}{\textbf{231}} & \textcolor{orange}{\underline{60}} & \textcolor{red}{\textbf{132}} \\
ICLR         & 440 & 365 & 425 
             & 199 & 221 & 213 & 190 & 165 & 147 & \textcolor{orange}{\underline{242}} & 215 & \textcolor{orange}{\underline{232}} 
             & 155 & 272 & \textcolor{orange}{\underline{334}} & \textcolor{red}{\textbf{187}} & \textcolor{red}{\textbf{207}} & \textcolor{orange}{\underline{137}} 
             & \textcolor{orange}{\underline{78}} & 50 & \textcolor{orange}{\underline{67}} & 30 & 49 & 57 \\
ICML         & 193 & 183 & 177 
             & \textcolor{orange}{\underline{402}} & \textcolor{orange}{\underline{443}} & \textcolor{orange}{\underline{494}} & 228 & \textcolor{red}{\textbf{309}} & \textcolor{red}{\textbf{305}} & 190 & 231 & \textcolor{red}{\textbf{247}} 
             & 137 & 184 & 196 & \textcolor{orange}{\underline{163}} & 121 & 74 
             & 53 & 30 & 48 & 39 & 31 & 36 \\
\bottomrule
\end{tabular}
\label{tab:comprehensive-venue-errors}
\end{table*}

\begin{figure*}[t]
\centering
\begin{minipage}[b]{0.2\textwidth}
\centering
\small
\setlength{\tabcolsep}{3pt}
\renewcommand{\arraystretch}{0.95}
\begin{tabular}{@{}lr@{}}
\toprule
\textbf{Category} & \textbf{\%} \\
\midrule
Compound names & \textbf{27.5} \\
Typographical & 24.0 \\
Nickname & 23.8 \\
Transliteration & 7.0 \\
First name error & 5.7 \\
Last name error & 3.8 \\
Author missing & 3.6 \\
Wrong person & 1.7 \\
Deadname & 1.5 \\
Other & 1.2 \\
\bottomrule
\end{tabular}
\vspace{0.5em}

\textbf{(a)} Discrepancy types (n=2,738)
\end{minipage}%
\hspace{0.02\textwidth}%
\begin{minipage}[b]{0.2\textwidth}
\centering
\small
\setlength{\tabcolsep}{2pt}
\renewcommand{\arraystretch}{0.95}
\begin{tabular}{@{}lrrr@{}}
\toprule
\textbf{Subcategory} & \textbf{\%} & \textbf{First} & \textbf{Last} \\
\midrule
Multi-part & 61.1 & 315 & 146 \\
Hyphenated & 22.5 & 61 & 109 \\
Particle (de/van) & 6.5 & 3 & 46 \\
Title (Md.) & 6.5 & 47 & 2 \\
Apostrophe & 1.7 & 0 & 13 \\
Patronymic & 0.9 & 0 & 7 \\
Flipped & 0.7 & 4 & 1 \\
\bottomrule
\end{tabular}
\vspace{0.5em}

\textbf{(b)} Compound breakdown (n=754)
\end{minipage}%
\hspace{0.1\textwidth}%
\begin{minipage}[b]{0.4\textwidth}
\centering
\includegraphics[width=\textwidth]{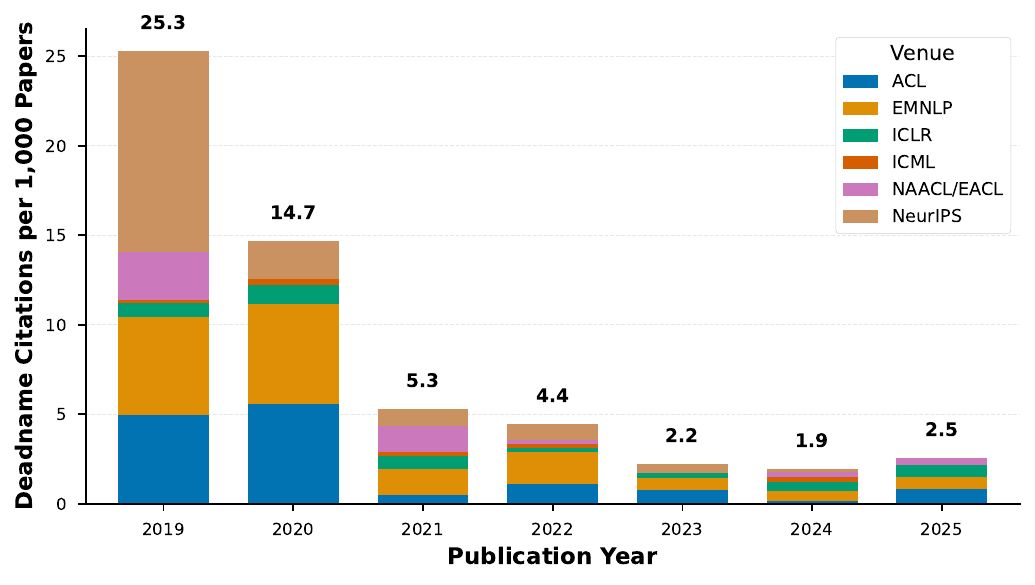}
\Description{Line chart showing deadname citations per 1,000 papers across six AI and NLP venues (ACL, EMNLP, ICLR, ICML, NAACL/EACL, NeurIPS) from 2019 to 2025.
The aggregate rate falls from 25.3 in 2019 to 14.7 in 2020, 5.3 in 2021, 4.4 in 2022, 2.2 in 2023, and 1.9 in 2024, before a slight increase to 2.5 in 2025.}
\vspace{-1.5em}

\textbf{(c)} Deadnames per 1,000 papers
\end{minipage}
\caption{Annotation study results. \textbf{(a)}~Distribution of name discrepancy types shows 75\% stem from compound names, typos, and nicknames. \textbf{(b)}~Compound name breakdown reveals multi-part names (61.1\%) and hyphenated surnames (22.5\%) dominate; First/Last columns show affected name component counts. \textbf{(c)}~Deadnaming rates dropped 92\%: From 25.3 (2019) to 1.9 (2024) per 1,000 papers.}
\label{fig:annotation-results}
\end{figure*}

\subsubsection{Annotation Study}
\label{sec:annotation}

Automated detection identifies \emph{that} names differ; understanding \emph{why} requires human judgment.
We uniformly sampled 3,000 discrepancies from the errors flagged by our reference checking pipeline.
Three annotators with training in naming conventions and name change policies independently classified each case.
Annotators viewed pairs of names: The \textcolor{blue}{cited name} alongside the \textcolor{teal}{authoritative name} from database records.
To protect author privacy, annotators never searched for individuals or attempted to identify specific researchers.
After excluding parsing artifacts and false positives, we retained 2,738 valid cases for analysis.
Annotators classified each discrepancy according to the typology in the previous section and also flagged instances of deadnaming (see Appendix \cref{ref:annotation-guidelines} for details), achieving strong inter-rater agreement ($\kappa = 0.86$ on 2738 instances).
Disagreements were resolved through discussion until consensus.
We then performed fine-grained annotation on the 754 compound name errors:

\begin{enumerate}
    \item \textbf{Multi-part names} (61.1\%): Names missing components 
    Example: \textcolor{blue}{M. Raghavan} $\rightarrow$ \textcolor{teal}{Anand Mohan Raghavan}.

    \item \textbf{Hyphenated surnames} (22.5\%): Hyphenated names split or partially dropped.
    Example: \textcolor{blue}{Garcia} $\rightarrow$ \textcolor{teal}{Garcia-Lopez}.

    \item \textbf{Name particles} (6.5\%): Particles like ``de,'' ``van,'' or ``von'' mishandled.
    Example: \textcolor{blue}{Berg} $\rightarrow$ \textcolor{teal}{van den Berg}.
    \item \textbf{Honorific titles} (6.5\%): Titles like ``Md.'' treated are parsed incorrectly.
    Example: \textcolor{blue}{M. Rahman} $\rightarrow$ \textcolor{teal}{Md. Aminur Rahman}.

    \item \textbf{Apostrophes} (1.7\%): Names with apostrophes incorrectly parsed.
    Example: \textcolor{blue}{OConnor} $\rightarrow$ \textcolor{teal}{O'Connor}.

    \item \textbf{Patronymics} (0.9\%): Patronymic elements like ``bin'' or ``binti'' dropped.
    Example: \textcolor{blue}{Hassan} $\rightarrow$ \textcolor{teal}{Ahmad bin Hassan}.

    \item \textbf{Name flipping} (0.7\%): First and last names swapped.
    Example: \textcolor{blue}{Madame Morrible} $\rightarrow$ \textcolor{teal}{Morrible Madame}.
\end{enumerate}

The annotation results are shown in Figure \ref{fig:annotation-results} which shows that citation errors happen predominantly due to compounded name errors.
Figure~\ref{fig:annotation-results}(c) shows the resulting temporal trends in deadnaming rates.

\subsection{Findings}
\label{sec:experiment}

\subsubsection{Venues with proactive name change policies had significantly lower error rates in citations.}

We compared citation error rates between venues with visible and accessible name change policies (ACL, EMNLP, NAACL/EACL, NeurIPS) and those without (AAAI, FAccT, ICLR, ICML) using papers published 2019--2025.
Venues with policies showed aggregate total 899 errors per 1,000 papers; venues without show 996.
This difference was statistically significant ($p < 0.001$; Cohen's $h = 0.34$),\footnote{We conducted a chi-square test on aggregate error counts ($\chi^2 = 93.97$, $p < 0.001$) and confirmed with a two-sample rate test ($z = -9.69$, $p < 0.001$).
The venue-level analysis uses a Mann-Whitney U test ($U = 2$), with reduced significance due to comparing only four venues per group rather than the absence of a meaningful effect.} with policy-lacking venues exhibiting approximately 10\% more citation errors.
Aggregate total numbers were high, as we observed that nearly every bibliography has a naming error. 
Venue-level analysis confirmed this pattern ($p = 0.057$), though reduced significance reflects limited power from comparing only four venues per group.
We reach a similar conclusion with citation normalization that venues with proactive name change policies exhibited significantly lower citation error rates ($\chi^2 = 32.81$, $p < 0.001$, see \cref{sec:errors}).
These findings showed that publication culture with proactive policies and supporting infrastructure reduced naming errors in citations.

\subsubsection{Deadnaming declined sharply after 2021.}
Rates dropped 92\% from 25.3 per 1,000 papers (2019) to 1.9 (2024).
NeurIPS improved the most, falling from 21.5 to 0.5 per 1,000.
ACL and EMNLP achieved comparable reductions despite higher starting rates (39.1\% and 29.6\% respectively).
We believe that this decline resulted from the policy efforts around this time: ACL developed citation checking practices in 2021 \cite{pranav2021naacl}, NeurIPS implemented its policy in 2021, and Queer in AI advocacy made citation practices a community priority~\citep{Queerinai_2023}.
Despite substantial improvement, deadnaming persisted in our data
Figure~\ref{fig:annotation-results}(c) shows a slight concerning uptick in the instances of deadnaming in 2025.

\subsubsection{Anglo-centric systems harm those with non-normative names.}
Researchers who do not fit naming norms in academia faced systematic citation errors: 75\% of discrepancies stem from compound names (27.5\%), typographical errors (24.0\%), and nicknames (23.8\%) (Figure~\ref{fig:annotation-results}a).
Multi-part names without delimiters accounted for 61.1\% of compound errors due to truncation (Figure~\ref{fig:annotation-results}b).
Finally, hyphenated multi-part surnames triggered last name errors in 64.1\% of cases.

\subsubsection{Annotators flagged fabricated citations.}
The ``wrong person'' category (1.7\%) captured citations that named entirely incorrect authors or referenced nonexistent papers.
These errors differed from outdated names; they suggest fabrication rather than propagation.
Such patterns were consistent with LLM-assisted writing, where models fabricate plausible-sounding but fictitious references.
\section{Recommendations}
\label{sec:recs}

Based on analyzing existing policies, surveying and interviewing scholars who have changed their names, and our empirical findings, we provide three recommendations for inclusive publishing based on policies, technical implementations, and community engagement.

\paragraph{Recommendation 1: Pro-Actively Implement Inclusive Publishing Policies.}
Effective name change policies must prioritize author autonomy while minimizing administrative overhead.
These policies should embody core principles of \textit{accessibility, comprehensiveness, invisibility, expediency, and systematic maintenance} \cite{tanenbaum2021vision}.
Publishers should implement immediate name changes upon request without requiring legal documentation.
They should ensure synchronization across all digital artifacts and maintain automated monitoring to prevent recurrence of outdated names.
Conferences must move beyond symbolic policy adoption to active implementation and enforcement.
Every call for papers should include name change policy information in its authorship section.
Conferences should require authors to register ORCID identifiers during submission, enabling consistent author identification across venues.
Since survey data shows that 50\% of respondents were unsure whether publishers in their field have formal policies, and multiple participants learned about name change options only through informal networks, Discord servers, or chance conversations with senior colleagues, documentation and discoverability should be a key priority.
Every published proceeding should prominently display instructions for reporting errors and requesting corrections.
Publishers should also display name change policies prominently in author guidelines, submission systems, and help documentation.

\paragraph{Recommendation 2: Improve Publishing Infrastructure.}
Current policies do not extend to updating citations in already-published articles.
We recommend that every publisher integrate automated name checking of references into their submission workflows.
Our analysis demonstrates the feasibility of this approach: We parsed over 100,000 papers, validated author names against authoritative databases, and identified systematic patterns of naming errors.
Such pipelines catch errors before publication, preventing future circulation of incorrect names in the scholarly record.
Automated verification can flag discrepancies without exposing specific name changes to submitting authors, preserving the privacy of cited authors at scale.
To support adoption, we release an open-source reference checker that authors and publishers can run on submitted bibliographies.\footnote{\url{https://github.com/pranav-ust/cite-updater}}
The tool parses references from a PDF, queries authoritative databases, and reports name discrepancies for the user to review before submission.
Further in the future, publishers should transition from static PDFs to HTML-based publishing with dynamic name rendering.
Multiple interview participants envisioned systems where author names are ``\textit{pulled down from ORCID and rendered}'' at display time rather than hard-coded in documents.
P7 described the ideal: ``\textit{If you update your name on your ORCID, it updates your name in all of the places that you've published\ldots That would be a magical, beautiful world.}''
This architecture treats ORCID as the authoritative presentation layer, enabling a single update to propagate across all publications.
HTML-based publishing would also improve accessibility for disabled users, addressing multiple justice concerns simultaneously~\citep{Kumar2024A11Y}.

\paragraph{Recommendation 3: Engage and Develop Policies with Affinity Groups.} 
Trans and queer scholars initiated and led the name change policy movement in academic publishing.
Early advocates faced significant resistance, including harassment severe enough to collapse initial working groups.
Despite these barriers, sustained advocacy by Queer in AI and the Name Change Policy Working Group led to policy adoption across major AI venues \cite{Queerinai2023}.
These policies benefit all scholars.
Several cisgender participants in our study learned about name change processes through trans advocacy, demonstrating how infrastructure improvements driven by marginalized groups create broader impact.
We encourage venues to engage with and support advocacy groups to build more inclusive scholarly communities.

\section{Conclusion}
In this paper, we advocate for inclusive naming and name change policies in academic publishing and provide concrete tools and recommendations for implementation.
This paper demonstrates how policies led by marginalized researchers have transformed the publishing landscape, proving that equitable infrastructure is not a niche concern, but a fundamental benefit to the scholarly ecosystem as a whole.
Our empirical analysis shows that venues with proactive visible policies have significantly fewer citation errors, though error rates persist even in venues with policies.
Our interviews reveal the extensive emotional labor and advocacy work required to achieve current policies, as well as ongoing challenges with current citation propagation methods, search engine delays, and information gaps. 
These failures reflect a denial of data sovereignty and researchers should hold authority over their own scholarly records.
We hope this work enables scholars to focus on advancing knowledge rather than fighting infrastructure.
\textbf{No researcher should expend countless hours pursuing a basic human dignity: To be properly named.}

\section*{Acknowledgements}
This paper would not have been possible without the trans activists and volunteers of the Name Change Policy Working Group and Queer in AI, whose advocacy established the policies analyzed in this paper and whose work informs our recommendations.
We are grateful to our survey respondents and interview participants for sharing their experiences with academic naming systems.
We also like to thank Simon Laatz and Fynn Wyroslawski in helping us out with annotations.
The first and last author are funded under the Excellence Strategy of the German Federal Government and States.

\section*{Ethical Considerations}
Our survey and interviews with researchers are experiments with human subjects. 
All data collection procedures and interview protocols here were approved by from the Ethics Committee of the University of Hamburg Business School, and we obtained informed consent from all participants for participation in the survey and interviews.
Interview participants gave consent for direct quotes and all other information in this publication, and are referenced with participant IDs.
The citation error analysis also involves sensitive data about researchers' names and identities, with particularly serious safety risks for transgender researchers.
Annotations were performed by researchers who were trained about issues with name changes.
Annotators were explicitly instructed not to search for authors' names during the annotation process to protect privacy.

\section*{Author Contributions}
We describe author contributions following the CRediT taxonomy.
\textbf{A Pranav}: Conceptualization, Investigation, Methodology, Writing - Original Draft.
\textbf{Vagrant Gautam}: Investigation, Mentorship, Writing - Review \& Editing.
\textbf{Martin Mundt}, \textbf{Jordan Taylor}, \textbf{William Agnew}, \textbf{Franziska Sofia Hafner}, \textbf{Daniel Chechelnitsky}, \textbf{Arjun Subramonian}, and \textbf{Anne Lauscher}: Writing - Review \& Editing.

\section*{Competing Interests}
Several authors are members of, or have collaborated with, advocacy groups referenced in this paper, including Queer in AI and the Name Change Policy Working Group.
The authors declare no financial competing interests.

\section*{Positionality Statement}
This research was led by authors who have changed their names in academia for various reasons, and was conducted in collaboration with queer authors and researchers.
Several authors have participated in name change policy advocacy at venues analyzed in this study.
Our positions as researchers personally affected by academic naming infrastructure shaped our research questions, our methodological choices, and our interpretation of findings.


\section*{Generative AI Usage Statement}

No AI tools were used in the preparation of the manuscript.





\newpage

\bibliographystyle{ACM-Reference-Format}
\bibliography{sample-base}

\appendix

\section{Survey Details}
\label{app:survey}

The full survey (apart from frontmatter to obtain consent and endmatter to obtain contact details for participants interested in being interviewed) is provided below.
All questions are optional.

\begin{enumerate}
    \item[] \textbf{Preliminary details}
    \item What’s your primary discipline / area of study? \\
    \texttt{(text box)}
    \item What was the primary reason for changing (or considering changing) your name? \\
    $\square$ Gender transition \\
    $\square$ Marriage \\
    $\square$ Divorce \\
    $\square$ Cultural or religious reasons \\
    $\square$ Academic visibility \\
    $\square$ Immigration \\
    $\square$ Other \texttt{(text box)}
    \item Which components of your name have changed or will change? (choose one or explain in the other) \\
    $\square$ First/given name (e.g., \textit{Jane Smith} $\rightarrow$ \textit{John Smith}, \textit{J. Smith} $\rightarrow$ \textit{J.K. Smith}) \\
    $\square$ Middle name (e.g., adding, removing, or changing middle names) \\
    $\square$ Last/family/surname (taking, adding or dropping a last/family name, e.g., \textit{Jae Smith} $\rightarrow$ \textit{Jae Jones}, \textit{Jae Smith} $\rightarrow$ \textit{Jae Smith-Jones}) \\
    $\square$ Full name change (changing all components of your name) \\
    $\square$ Name ordering (e.g., \textit{Joe Sith} $\rightarrow$ \textit{Sith Joe}) \\
    $\square$ Cultural name elements (e.g., adding / removing / reformatting / hyphenating particles like "\textit{van}," "\textit{de}," "\textit{bin}" or diacritics like \textit{Janiça}) \\
    $\square$ Other \texttt{(text box)}
    \item What is your primary method for finding citations when writing papers? (select your most frequently used method) \\
    $\odot$ Publisher website (conference/journal) \\
    $\odot$ Google Scholar \\
    $\odot$ Semantic Scholar \\
    $\odot$ arXiv \\
    $\odot$ Reference manager (Mendeley, Zotero, etc.) or Online BibTeX generators \\
    $\odot$ Copy from other papers \\
    $\odot$ Manual \\
    $\odot$ Other \texttt{(text box)}
    \item Where do you primarily publish? (list up to 5 main venues/journals) \texttt{(text box)} \\
    \item[] \textbf{In which contexts have you changed or considered changing your name?}
    \item Changing names socially: Using a new name with friends, family, colleagues, or in daily life (e.g., introductions, email signatures, social media) \\
    $\odot$ Yes \\
    $\odot$ No \\
    $\odot$ Considering in future \\
    $\odot$ Other \texttt{(text box)}
    \item Changing names academically: Changing your name on publications, conference presentations, academic profiles, or institutional records (e.g., updating published papers, ORCID, Google Scholar) \\
    $\odot$ Yes \\
    $\odot$ No \\
    $\odot$ Considering in future \\
    $\odot$ Other \texttt{(text box)} \\
    \item[] \textbf{Have you encountered challenges when requesting name changes from publishers or venues?}
    \item Please list any publishers/venues where you faced difficulties and describe your experience: \texttt{(text box)} \\
    \item[] \textbf{Academic Impact}
    \item How often are you still cited using your previous name in recent publications? \\
    $\odot$ Never (0\%) \\
    $\odot$ Rarely (1-20\%) \\
    $\odot$ Sometimes (21-40\%) \\
    $\odot$ Often (41-60\%) \\
    $\odot$ Very often (61-80\%) \\
    $\odot$ Almost always (81-100\%) \\
    $\odot$ I haven't tracked this
    \item When publishers do update your name, how thoroughly is your previous name removed? \\
    $\square$ Complete removal - previous name no longer visible anywhere \\
    $\square$ Partial update - some records updated but not others (e.g., metadata changed but not PDFs) \\
    $\square$ Visible correction - both names shown (e.g., "formerly published as...") \\
    $\square$ Erratum/corrigendum - change noted as a "correction" implying error \\
    $\square$ Other
    \item The process of correcting my name in academic publications has impacted me in the following ways: \\
    $\square$ Fragmented citation metrics affecting career evaluation (split h-index, citation counts) \\
    $\square$ Unwanted disclosure of my trans/personal status to colleagues or public \\
    $\square$ Caused significant stress or anxiety \\
    $\square$ Required substantial time away from research/teaching \\
    $\square$ Led me to avoid citing my own previous work \\
    $\square$ Made me reconsider my academic career \\
    $\square$ Affected my sense of belonging in my field \\
    $\square$ Positive experiences with supportive publishers \\
    $\square$ Other
    \item Do the primary publishers/venues in your field have formal name change policies? \\
    $\odot$ All or most have clear, well-implemented policies: I can understand how to change my name and it is implemented quickly and discreetly \\
    $\odot$ All or most have policies, but implementation is inconsistent \\
    $\odot$ Mixed - some have good policies, others don't \\
    $\odot$ Few or none have formal policies \\
    $\odot$ I'm not sure / haven't checked
    \item Optional: Name specific publishers that you found struggled with good name change policies
    \item Is there anything else about your experience with name changes as a researcher that you'd like to share?
\end{enumerate}

\section{Interview Details}
\label{app:interview}

We ran semi-structured interviews around the following questions, all of which were optional:

\begin{enumerate}
    \item Can you tell us a bit about why you changed your name and how you decided on the new name? When did you change the name?
    \item Can you tell us more about your experiences of changing your name at different venues---we want to know about pain points as well as successes? (Follow-up: How has this propagated, e.g., to search engines such as google scholar, etc., or in LLM-generated content)
    \item How has your name change impacted you in terms of your mental health and your career---both during the process and after it was done?
    \item Where did you find the information about how you can change your name in academia?
    \item What changes do you want to see in how we handle name changes in academia?
    \item Is there anything else you want to share with us that we have not already covered?
\end{enumerate}

\section{Training the Reference Parser}
\label{sec:parser}

\subsection{Model Architecture}
We built our reference parser on top of BERT-Base, augmented with a Conditional Random Field (CRF) layer for sequence labeling. We chose this architecture over more complex models like RoBERTa or GPT variants because it offers a good balance between computational efficiency and accuracy for our specialized task.

\subsection{Dataset Construction}
To create a training dataset, we collected 3,000 papers published between 2014--2024 from major computer science conferences:
\begin{itemize}
    \item 500 papers from ACL anthology
    \item 500 papers from ICML
    \item 500 papers from ICLR
    \item 500 papers from AAAI
    \item 500 papers from NeurIPS
    \item 500 papers from FAccT
\end{itemize}

For each paper, we obtained both the PDF and its corresponding BibTeX entries from arXiv. We developed a pipeline to automatically extract and align references from PDFs with their corresponding BibTeX entries. 
The dataset was preprocessed and formatted according to GROBID's training data specifications, which involve labeling different components of bibliographic references using a specialized XML schema. Each reference was annotated with the following components:
\begin{itemize} 
    \item Author names (with separate tags for first and last names)
    \item Paper titles
    \item Venue names
    \item Publication years
    \item Volume and issue numbers
    \item Page numbers
    \item DOI identifiers
\end{itemize}

To ensure data quality, we performed several cleaning steps:
\begin{itemize} 
    \item Removed references with missing or incomplete BibTeX entries
    \item Standardized venue names using a mapping table
    \item Validated DOI identifiers for accuracy
\end{itemize}

\subsection{Training Setup}
The model was trained on an NVIDIA RTX 2080 GPU with the following hyperparameters:
\begin{itemize} 
    \item Number of epochs: 4
    \item Learning rate: 1e-4
    \item Batch size: 32
    \item Optimizer: AdamW
    \item Weight decay: 0.01
\end{itemize}

We used a linear learning rate schedule with warmup over the first 10\% of training steps.

\subsection{Evaluation}

Similarly, we take a random sample of 300 papers and construct our test set.
These results represent an improvement over the base GROBID model, which achieves a macro-averaged field based (average F1 score of all the fields like authors, title, DOI, journals) F1 score of 0.96 on our test set. 

\section{Annotation Process}
\label{ref:annotation-guidelines}

Three annotators with training in name changing practices in academia
independently analyzed citation errors in the dataset. 
Given the delicate nature of dealing with deadnames, annotators signed confidentiality agreements and followed strict privacy protocols.
The institutional ethics board approved the annotation process.
The annotators achieved an inter-annotator agreement of $\kappa = 0.86$, indicating strong reliability in their classifications. Each citation was evaluated according to the following guidelines:

\begin{enumerate}
    \item \textbf{First Name Changes}
    \begin{itemize}
        \item Identify significant changes in first names (e.g., ``Jon'' vs ``Jonathan'')
        \item Do not flag abbreviated forms unless significantly different
    \end{itemize}
    
    \item \textbf{Last Name Changes}
    \begin{itemize}
        \item Mark substantial changes in last names (e.g., ``Smith'' vs ``Brown'')
        \item Note major structural changes like addition/removal of hyphenation
        \item Ignore minor spelling variations or typographical errors
        \item Do not investigate reasons for name changes
    \end{itemize}
    
    \item \textbf{Deadname Detection}
    \begin{itemize}
        \item Mark entries that match known deadname patterns, which are mainly different first names starting with the same initial and the same last name
        \item Do not search for or verify author information
        \item Maintain absolute confidentiality of identified deadnames
        \item Skip entry if uncertain rather than investigating further
    \end{itemize}
    
    \item \textbf{Parsing Errors}
    \begin{itemize}
        \item Check if citation was parsed correctly from the PDF
        \item Identify if author names were correctly extracted from citation string
        \item Note if parsing resulted in incorrect author attribution
        \item Flag cases where parsing led to missing or merged author names
    \end{itemize}
\end{enumerate}

\noindent\textbf{General Instructions:}
\begin{itemize} 
    \item Do not look up names in any external databases
    \item Never search for author information online
    \item Keep all identified deadnames strictly confidential
    \item Work independently before comparing results
    \item Flag uncertain cases rather than making assumptions
    \item Focus only on significant name changes, not minor variations
\end{itemize}

\section{Policy Overview}
\label{sec:policy}

\begin{figure}[h]
\centering
\includegraphics[width=\columnwidth]{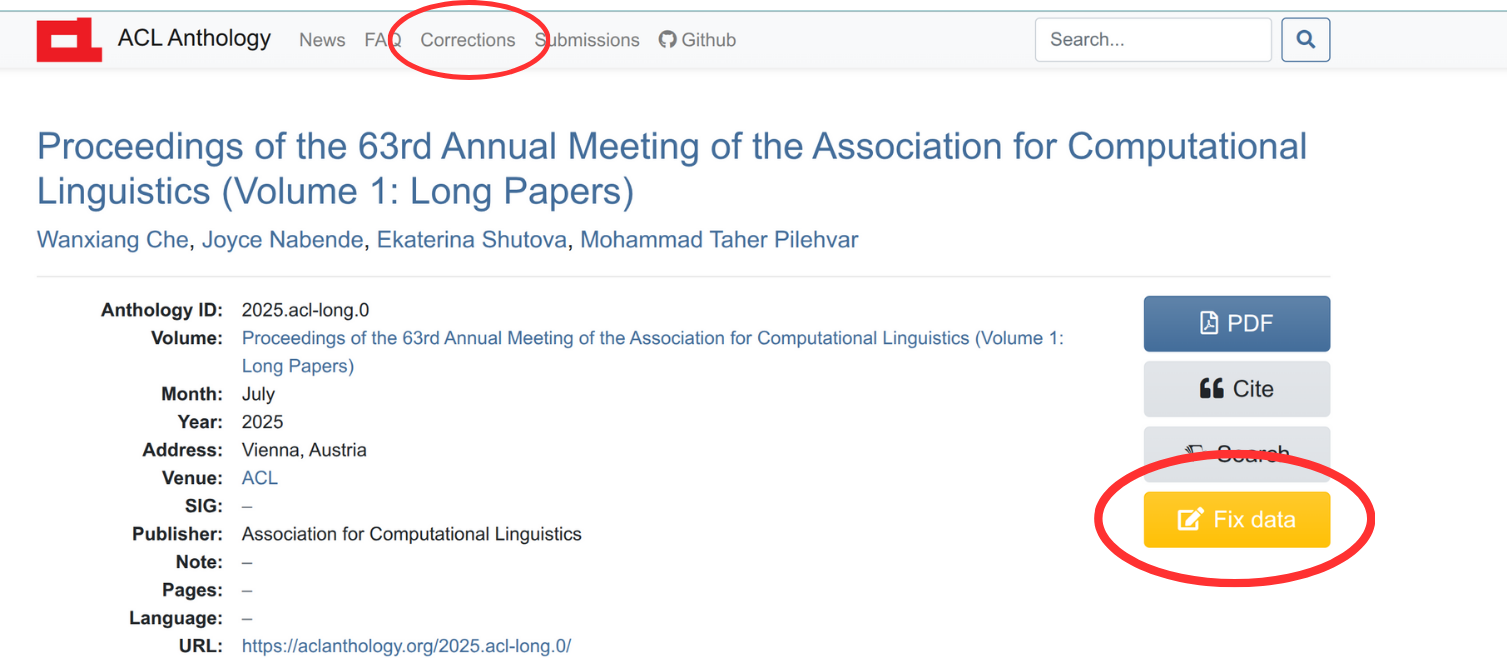}
\caption{Visible and accessible option for authors to correct and fix data of their work in ACL proceedings.}
\label{fig:acl}
\Description{Screenshot of an ACL Anthology proceedings page for the 63rd Annual Meeting of the Association for Computational Linguistics, Volume 1: Long Papers (ACL 2025).
Two interface elements are circled in red: the ``Corrections'' link in the top navigation bar, and a yellow ``Fix data'' button on the right side of the page.}
\end{figure}

\begin{figure}[h]
\centering
\includegraphics[width=\columnwidth]{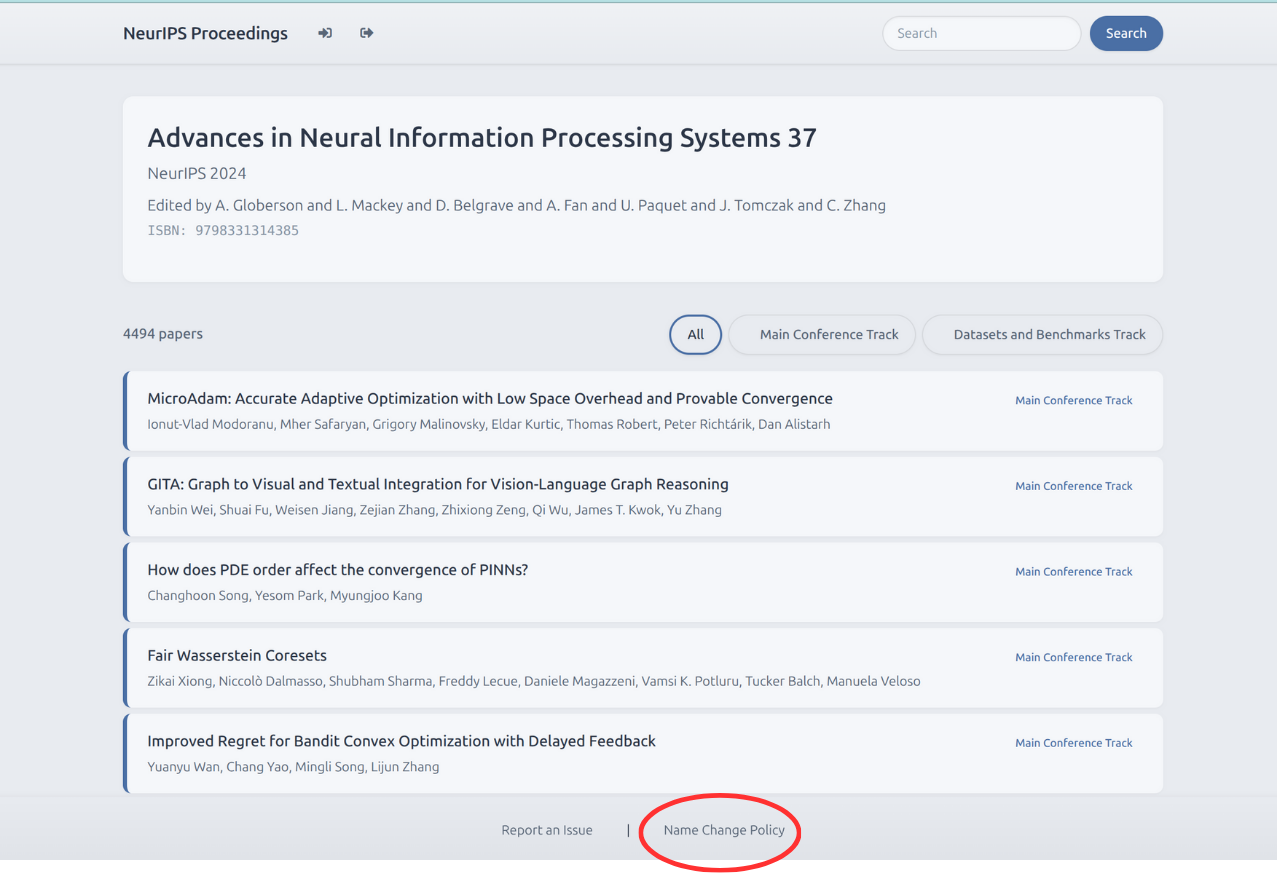}
\caption{Name Change Policy visible in every proceeding of NeurIPS.}
\Description{Screenshot of the NeurIPS Proceedings website showing the landing page for Advances in Neural Information Processing Systems 37 (NeurIPS 2024), listing 4,494 papers.
A red circle highlights the ``Name Change Policy'' link in the page footer, alongside a ``Report an Issue'' link.}
\label{fig:neurips}
\end{figure}

We provide a comprehensive overview of name change policies across major computer science conferences as of January 2026 in Table \ref{tab:name-change-policies}. The table documents policies and documentation across different publishers and venues. 

\paragraph{ACL Anthology.} ACL instituted the name change policy in 2023. Besides this, ACL anthology deals with names changes discreetly. Every proceeding of ACL has a visible option of fixing data, which is shown in Figure \ref{fig:acl}.
Researchers can also submit pull requests for their fixes.
ACL also provides high quality updated BibTeX options for every proceeding. This aggregated bib file is also present in CfP and templates of ACL. 

\paragraph{NeurIPS.} NeurIPS instituted name change policy in 2022. Every proceeding of NeurIPS has a visible link to name change policy and reporting issue link, see Figure \ref{fig:neurips}. Similarly with ACL, NeurIPS provides dynamic bibtex citations for every proceedings which are immediately fixed after the name updates.

\paragraph{FAccT.} While ACM has a name change policy, FAccT has not made this visible on their website. There are no visible options to fix the data like NeurIPS and ACL does.

\begin{table*}[htbp]
\small
\caption{Name change policies in the major computer science conferences.}
\centering
\renewcommand{\arraystretch}{1.2}
\begin{tabular}{p{0.5\textwidth}p{0.2\textwidth}p{0.2\textwidth}}
\toprule
\textbf{Conference} & \textbf{Publisher} & \textbf{Name Change Policy} \\
\midrule
\rowcolor{gray!10}
AAAI (National Conference of the American Association for Artificial Intelligence) & AAAI Press & No policy available \\
AAMAS (International Conference on Autonomous Agents and Multiagent Systems) & ACM & Yes (\href{https://www.acm.org/publications/policies/author-name-changes}{ACM Policy}) \\
\rowcolor{gray!10}
ACL (Association for Computational Linguistics) & ACL Anthology & Yes (\href{https://www.aclweb.org/adminwiki/index.php/ACL_Name_Change_Policy}{ACL Policy}) \\
ACMMM (ACM Multimedia Conference) & ACM & Yes (\href{https://www.acm.org/publications/policies/author-name-changes}{ACM Policy}) \\
\rowcolor{gray!10}
ASPLOS (Architectural Support for Programming Languages and Operating Systems) & ACM & Yes (\href{https://www.acm.org/publications/policies/author-name-changes}{ACM Policy}) \\
CAV (Computer Aided Verification) & Springer & Yes \\
CHI (International Conference on Human Factors in Computing Systems) & ACM & Yes (\href{https://www.acm.org/publications/policies/author-name-changes}{ACM Policy}) \\
\rowcolor{gray!10}
COLT (Annual Conference on Computational Learning Theory) & PMLR & No policy available \\
CRYPTO (International Cryptology Conference) & Springer & Yes \\
\rowcolor{gray!10}
CVPR (Conference on Computer Vision and Pattern Recognition) & IEEE & Yes (\href{https://journals.ieeeauthorcenter.ieee.org/become-an-ieee-journal-author/publishing-ethics/guidelines-and-policies/ieee-author-name-change-policy/}{IEEE Policy}) \\
EMNLP (Conference on Empirical Methods in Natural Language Processing) & ACL Anthology & Yes (\href{https://www.aclweb.org/adminwiki/index.php/ACL_Name_Change_Policy}{ACL Policy}) \\
\rowcolor{gray!10}
FAccT (Fairness, Accountability, and Transparency) & ACM & Yes (\href{https://www.acm.org/publications/policies/author-name-changes}{ACM Policy}) \\
HPCA (IEEE Symposium on High Performance Computer Architecture) & IEEE & Yes (\href{https://journals.ieeeauthorcenter.ieee.org/become-an-ieee-journal-author/publishing-ethics/guidelines-and-policies/ieee-author-name-change-policy/}{IEEE Policy}) \\
\rowcolor{gray!10}
ICDE (IEEE International Conference on Data Engineering) & IEEE & Yes (\href{https://journals.ieeeauthorcenter.ieee.org/become-an-ieee-journal-author/publishing-ethics/guidelines-and-policies/ieee-author-name-change-policy/}{IEEE Policy}) \\
ICDM (IEEE International Conference on Data Mining) & IEEE & Yes (\href{https://journals.ieeeauthorcenter.ieee.org/become-an-ieee-journal-author/publishing-ethics/guidelines-and-policies/ieee-author-name-change-policy/}{IEEE Policy}) \\
\rowcolor{gray!10}
ICLR (International Conference on Learning Representations) & OpenReview & No policy available \\
ICML (International Conference on Machine Learning) & PMLR & No policy available \\
\rowcolor{gray!10}
ICSE (International Conference on Software Engineering) & IEEE & Yes (\href{https://journals.ieeeauthorcenter.ieee.org/become-an-ieee-journal-author/publishing-ethics/guidelines-and-policies/ieee-author-name-change-policy/}{IEEE Policy}) \\
IJCAI (International Joint Conference on Artificial Intelligence) & AAAI Press & No policy available \\
\rowcolor{gray!10}
INFOCOM (IEEE Conference on Computer Communications) & IEEE & Yes (\href{https://journals.ieeeauthorcenter.ieee.org/become-an-ieee-journal-author/publishing-ethics/guidelines-and-policies/ieee-author-name-change-policy/}{IEEE Policy}) \\
ISCA (International Symposium on Computer Architecture) & ACM & Yes (\href{https://www.acm.org/publications/policies/author-name-changes}{ACM Policy}) \\
\rowcolor{gray!10}
NDSS (Network and Distributed System Security Symposium) & Internet Society & No policy available \\
NeurIPS (Neural Information Processing Systems) & PMLR & Yes (\href{https://papers.nips.cc/}{NeurIPS Policy}) \\
\rowcolor{gray!10}
OSDI (USENIX Symposium on Operating Systems Design and Implementation) & USENIX & Yes (\href{https://www.usenix.org/name-change-policy}{USENIX Policy}) \\
PLDI (Programming Language Design and Implementation) & ACM & Yes (\href{https://www.acm.org/publications/policies/author-name-changes}{ACM Policy}) \\
\rowcolor{gray!10}
SIGCOMM (ACM Conference on Data Communication) & ACM & Yes (\href{https://www.acm.org/publications/policies/author-name-changes}{ACM Policy}) \\
SIGGRAPH (International Conference on Computer Graphics and Interactive Techniques) & ACM & Yes (\href{https://www.acm.org/publications/policies/author-name-changes}{ACM Policy}) \\
\rowcolor{gray!10}
SIGIR (International Conference on Research and Development in Information Retrieval) & ACM & Yes (\href{https://www.acm.org/publications/policies/author-name-changes}{ACM Policy}) \\
SIGMOD (ACM Special Interest Group on Management of Data Conference) & ACM & Yes (\href{https://www.acm.org/publications/policies/author-name-changes}{ACM Policy}) \\
\rowcolor{gray!10}
SOSP (ACM Symposium on Operating Systems Principles) & ACM & Yes (\href{https://www.acm.org/publications/policies/author-name-changes}{ACM Policy}) \\
STOC (ACM Symposium on Theory of Computing) & ACM & Yes (\href{https://www.acm.org/publications/policies/author-name-changes}{ACM Policy}) \\
\rowcolor{gray!10}
VLDB (International Conference on Very Large Databases) & PVLDB & No policy available \\
WWW (The Web Conference) & ACM & Yes (\href{https://www.acm.org/publications/policies/author-name-changes}{ACM Policy}) \\
\bottomrule
\end{tabular}
\label{tab:name-change-policies}
\end{table*}

\section{Alternative results on number of citations normalization}
\label{sec:errors}

An alternative metric to our per 1000 paper metric would be citation errors per 1000 citations. Following are the results of the errors per 1000 citations.
Venues with policies showed 17.1 errors per 1,000 citations, compared to 18.1 errors per 1,000 citations for venues without policies.
A chi-square test revealed this difference was statistically significant ($\chi^2 = 32.81$, $p < 0.001$).
A two-sample proportion test confirmed the result ($z = -5.73$, $p < 0.001$).
However, the effect size (Cohen's $h = 0.008$) indicates a negligible practical difference at the citation level.
Venues without name change policies exhibited 1.06 times the error rate of venues with policies, representing approximately 6\% more citation errors.

At the venue level, a Mann-Whitney U test comparing aggregate error rates across the eight venues showed no significant difference ($U = 4$, $p = 0.171$, one-tailed).
A permutation test with 10,000 iterations confirmed this result ($p = 0.155$).
The reduced significance at the venue level reflects both limited statistical power from comparing only four venues per group and greater variance in per-venue rates.

\begin{table*}[t]
\centering
\caption{Errors per 1,000 citations in paper bibliographies (2019--2024).
Citation-level analysis reveals similar patterns to paper-level analysis: Venues without name change policies exhibit higher compound-name error rates.
\textcolor{red}{\textbf{Red}} = highest; \textcolor{orange}{\underline{orange}} = second.
Typo = typographical; LN-C/FN-C = last/first name compound; Nick = nickname; LN-E = last name error; Trans = transliteration; AO = author order; WP = wrong person.
'19 = 2019--21; '22 = 2022--23; '24 = 2024--25.}
\footnotesize
\renewcommand{\arraystretch}{1.15}
\setlength{\tabcolsep}{2.8pt}
\begin{tabular}{@{}l*{24}{r}@{}}
\toprule
& \multicolumn{3}{c}{\textbf{Typo}} 
& \multicolumn{3}{c}{\textbf{LN-C}} 
& \multicolumn{3}{c}{\textbf{FN-C}} 
& \multicolumn{3}{c}{\textbf{Nick}} 
& \multicolumn{3}{c}{\textbf{LN-E}} 
& \multicolumn{3}{c}{\textbf{Trans}} 
& \multicolumn{3}{c}{\textbf{AO}} 
& \multicolumn{3}{c}{\textbf{WP}} \\
\cmidrule(lr){2-4} \cmidrule(lr){5-7} \cmidrule(lr){8-10} \cmidrule(lr){11-13} 
\cmidrule(lr){14-16} \cmidrule(lr){17-19} \cmidrule(lr){20-22} \cmidrule(lr){23-25}
\textbf{Venue} & \textbf{'19} & \textbf{'22} & \textbf{'24} 
& \textbf{'19} & \textbf{'22} & \textbf{'24} 
& \textbf{'19} & \textbf{'22} & \textbf{'24} 
& \textbf{'19} & \textbf{'22} & \textbf{'24} 
& \textbf{'19} & \textbf{'22} & \textbf{'24} 
& \textbf{'19} & \textbf{'22} & \textbf{'24} 
& \textbf{'19} & \textbf{'22} & \textbf{'24} 
& \textbf{'19} & \textbf{'22} & \textbf{'24} \\
\midrule
\multicolumn{25}{@{}l}{\cellcolor{gray!8}\textit{Venues with name change policies}} \\[2pt]
ACL          & 6.6 & 4.8 & 4.3 
             & 1.5 & 1.9 & 1.6 & 1.3 & 1.0 & 1.3 & 1.9 & 1.6 & 1.7 
             & 3.9 & 3.6 & 3.2 & 1.9 & 0.9 & 1.0 
             & 0.7 & \textcolor{red}{\textbf{0.7}} & 0.6 & 0.6 & 0.6 & 0.8 \\
EMNLP        & 6.5 & 4.4 & \textcolor{orange}{\underline{4.7}} 
             & 2.3 & 2.1 & 1.8 & 1.7 & 1.3 & 1.2 & 1.7 & 1.9 & 1.4 
             & \textcolor{orange}{\underline{4.4}} & 3.5 & 3.4 & 1.1 & 1.0 & 1.0 
             & 0.3 & 0.5 & 0.7 & 0.7 & 0.5 & \textcolor{orange}{\underline{0.8}} \\
NAACL/EACL   & \textcolor{red}{\textbf{7.0}} & \textcolor{red}{\textbf{5.0}} & 4.6 
             & 1.7 & 2.1 & 2.2 & 1.4 & 1.5 & 1.6 & 2.1 & 1.8 & 1.7 
             & 4.2 & \textcolor{red}{\textbf{4.8}} & \textcolor{orange}{\underline{3.6}} & 1.4 & 1.0 & 1.0 
             & 0.3 & 0.4 & 0.6 & 0.4 & \textcolor{red}{\textbf{0.8}} & 0.5 \\
NeurIPS      & 5.9 & 3.7 & 3.6 
             & 3.2 & 3.3 & 2.9 & \textcolor{orange}{\underline{2.9}} & 2.1 & 2.0 & 2.6 & 2.4 & 2.0 
             & 3.3 & 2.8 & 3.2 & \textcolor{orange}{\underline{1.9}} & 1.4 & 1.0 
             & 0.7 & 0.5 & 0.6 & 0.4 & 0.6 & 0.5 \\
\midrule
\multicolumn{25}{@{}l}{\cellcolor{gray!8}\textit{Venues without name change policies}} \\[2pt]
AAAI         & 1.9 & 1.8 & 1.9 
             & \textcolor{orange}{\underline{4.7}} & \textcolor{red}{\textbf{5.8}} & \textcolor{red}{\textbf{6.8}} & \textcolor{orange}{\underline{3.4}} & \textcolor{red}{\textbf{3.9}} & \textcolor{red}{\textbf{3.3}} & 2.9 & \textcolor{red}{\textbf{3.1}} & \textcolor{red}{\textbf{2.7}} 
             & 2.5 & 2.0 & 3.0 & 1.2 & 1.3 & 1.2 
             & 0.6 & 0.6 & \textcolor{orange}{\underline{0.8}} & 0.3 & 0.5 & 0.5 \\
FAccT        & 2.9 & 2.7 & 3.7 
             & \textcolor{red}{\textbf{3.9}} & 3.5 & 1.9 & 1.6 & 1.5 & 1.3 & 2.6 & 2.3 & 1.6 
             & 2.2 & \textcolor{orange}{\underline{3.7}} & 3.1 & 1.0 & \textcolor{orange}{\underline{1.5}} & \textcolor{red}{\textbf{1.6}} 
             & \textcolor{red}{\textbf{1.0}} & 0.2 & 0.7 & \textcolor{red}{\textbf{1.9}} & 0.4 & \textcolor{red}{\textbf{1.0}} \\
ICLR         & \textcolor{orange}{\underline{6.2}} & \textcolor{orange}{\underline{4.4}} & 4.6 
             & 2.7 & 2.7 & 2.3 & 2.4 & 2.0 & 1.6 & \textcolor{red}{\textbf{3.1}} & \textcolor{orange}{\underline{2.6}} & \textcolor{orange}{\underline{2.4}} 
             & 2.6 & 3.2 & \textcolor{red}{\textbf{3.8}} & \textcolor{red}{\textbf{2.4}} & \textcolor{red}{\textbf{2.5}} & \textcolor{orange}{\underline{1.5}} 
             & \textcolor{orange}{\underline{0.9}} & 0.5 & \textcolor{red}{\textbf{0.8}} & 0.5 & 0.6 & 0.6 \\
ICML         & 2.4 & 2.0 & 1.8 
             & \textcolor{red}{\textbf{5.2}} & \textcolor{orange}{\underline{4.9}} & \textcolor{orange}{\underline{5.0}} & \textcolor{red}{\textbf{3.5}} & \textcolor{orange}{\underline{3.4}} & \textcolor{orange}{\underline{3.1}} & \textcolor{orange}{\underline{3.1}} & 2.5 & 2.5 
             & 2.5 & 2.0 & 2.0 & 1.6 & 1.3 & 0.8 
             & \textcolor{orange}{\underline{0.9}} & 0.3 & 0.5 & 0.5 & 0.3 & 0.4 \\
\bottomrule
\end{tabular}
\label{tab:citation-errors}
\end{table*}

\end{document}